\documentclass[twocolumn,10pt]{asme2ej}

\usepackage{amssymb}
\usepackage{latexsym}

\usepackage{url}
\usepackage{xcolor}
\definecolor{Orange}{rgb}{.8,.349,.1}
\definecolor{Purple}{rgb}{.5, .0, .8}
\definecolor{Blue}{rgb}{0,0,180}
\definecolor{Green}{rgb}{0,100,0}

\usepackage[labelfont=bf, textfont=md]{caption}
\usepackage{mathtools}
\usepackage{subcaption}
\usepackage{url}

\usepackage{breakurl}
\newcommand{\AD}{$2\mathrm{d}_{0}$}
\newcommand{\BD}{$16\mathrm{d}_{0}$}
\newcommand{\CD}{$30\mathrm{d}_{0}$}
\newcommand{\Ad}{$2\mathrm{y}_{\mathrm{vo}}$}
\newcommand{\Bd}{$16\mathrm{y}_{\mathrm{vo}}$}
\newcommand{\Cd}{$30\mathrm{y}_{\mathrm{vo}}$}

\newcommand{\sfsix}{$\mathrm{SF}_{6}$}

\newcommand{\xRage}{xRAGE}
\newcommand{\simDD}{2D}
\newcommand{\simDDD}{3D}

\title{Validation methodologies for turbulent variable density flows: A jet case study \thanks{LA-UR-21-28743}}

\author{Austin Davis
    \affiliation{
	Department of Physics and Astronomy\\
	University of Victoria\\
	PO Box 3055 Victoria BC V8W 3P6\\
	Canada\\
    Email: adavis@uvic.ca
    }	
}

\author{Samuel Jones
    \affiliation{X Computational Physics\\ 
        Los Alamos National Lab\\
        Los Alamos, 87545\\
    }
}

\author{John J. Charonko
    \affiliation{Physics Division\\ 
        Los Alamos National Lab\\
        Los Alamos, 87545\\
    }
}

\author{Chris M. Malone
    \affiliation{X Computational Physics\\ 
        Los Alamos National Lab\\
        Los Alamos, 87545\\
    }
}

\author{Katherine Prestridge
    \affiliation{Physics Division\\ 
        Los Alamos National Lab\\
        Los Alamos, 87545\\
    }
}

\begin{document}

\maketitle

%%%%%%%%%%%%%%%%%%%%%%%%%%%%%%%%%%%%%%%%%%%%%%%%%%%%%%%%%%%%%%%%%%%%%%
\begin{abstract} % 150 words
{\it 
Comparisons studies between simulated variable density turbulent flows often consist of direct graphical representations where the level of agreement is determined by eye. This work demonstrates a formal validation methodology using an existing validation framework to examine the agreement between a simulated variable density jet flow and corresponding experimental data. Implicit large eddy simulations (ILES's) of a round jet and a plane jet with density ratio $s = 4.2$ were simulated using the compressible hydrodynamic code xRAGE. The jet growth, characterized by the spreading rates, was compared, and the difference between the simulations and the experiment was examined through jet structure diagnostics. The spreading rates were found to be larger than the experimental values, primarily due to resolution issues in the simulations, a fact that is quantified by the validation metric analysis.
}
\end{abstract}

%%%%%%%%%%%%%%%%%%%%%%%%%%%%%%%%%%%%%%%%%%%%%
\section{Introduction}
\label{sec:Intro}

% Intro
Implicit large eddy simulations (ILES's) of turbulent multi-physics flows are used in a variety of scientific applications such as the study of turbulent convection in the interior of stars \cite{Andrassy2020, Herwig2014, Cristini2019, Jones2016}, modeling the shock propagation in Type-II supernovae explosions \cite{Fields2020, Muller2017}, and in the study of turbulence \cite{Grinstein2021, Youngs2017, Karaca2012}, among others. Unlike large-eddy simulations (LES's), ILES's do not implement subgrid turbulence models and the behavior of the dissipation of the flow variables is left to the computational method \cite{Grinstein2007}.

Assessing the ability of ILES to accurately reproduce turbulent multi-physics flow requires high-quality experimental data of a known multi-physics flow so that a detailed validation study can be performed. The experimental data of Charonko and Prestridge \cite{Charonko2017} provides high spatial resolution, high sample count fields of density and velocity for air and \sfsix~jets in air coflow with quantified measurement uncertainties. The quality of the data and the associated uncertainties make this experimental data ideal for validation efforts focused on variable-density turbulent mixing. The variable density jet provides a good physical environment to test the turbulent mixing modeled by the simulation as the structure of the jet is governed primarily by the statistically stationary turbulent mixing within the shearing layer \cite[Ch. 5.1]{Pope2000}. 

%%%%
% P: History of jets 2D/3D
Comparisons of LES's to experimental variable density jets have been conducted in the past. Zhou et al. \cite{Zhou2001} computed LES's of heated buoyant jets in the near nozzle range that experience a pulsation instability, and compared their simulations to the experimental work of George et al. \cite{George1977} and Shabbir and George \cite{Shabbir1994}. Wang et al. \cite{Wang2008} used LES's with a dynamic Smagorinsky turbulence model to reproduce the experimental results of Djeridane et al. \cite{Djeridane1996} and Amielh et al. \cite{Amielh1996} with density ratios, $s = \rho_{\mathrm{jet}}/\rho_{\infty} = 0.14, 1.0, 1.52$. Foysi et al. \cite{Foysi2010} calculated plane and round jet LES's of heated jets with density ratios of $0.14, 1.0, 1.52$, and presented a density ratio dependent downstream velocity scaling, also comparing their results to Djeridane et al. \cite{Djeridane1996} and Amielh et al. \cite{Amielh1996}. Maragkos et al. \cite{Maragkos2014} computed LES's of a non-reacting $\mathrm{H}_{2}/\mathrm{CO}_{2}$ jet and compared to the experimental work of Smith et al. \cite{Smith1995} to show the influence of Reynolds number of differential diffusion. The experimental data, in this case, was limited and did not allow for a comparison of velocity or concentration fields. Although these studies, in general, show good agreement with the experimental data, quantified assessments of that agreement were not presented. Validation studies, rather than scaled, grid scale based comparisons, are required to determine the level of agreement of a particular jet model to the experimental measurements.

%%%%%%%%%%%%
% P: Validation methods
Validation workflows exist for testing a set of predicted model parameters against experimentally measured data \cite{Wilson2020, ASME2006, ASME2019, OberkampfRoy2010}. Wilson and Koskelo \cite{Wilson2020} outlines a validation workflow consisting of four steps. The \textit{Model Accuracy}, where the predictive accuracy is calculated with an appropriate validation metric. The \textit{Model Acceptability}, where the predictive accuracy of the model is compared to the model acceptance criteria. The \textit{Validation Evaluation}, where the details of the validation methodology, the intended use of the model, and acceptance criterion are examined with respect to the modeling applications. And finally, the \textit{Validation Recommendation}, where recommendations are offered to the modelers and users of the model based on the Validation Evaluation and the models' intended use. This workflow has been used in the past to assess the agreement between experimentally measured values and model predictions in inertial confinement fusion \cite{Wilson2018}.

Before a validation study can be conducted, appropriate model parameters need to be identified as characteristic of the physics under investigation. This study is concerned with variable density turbulent mixing, therefore to focus the study on mixing and limit the scope, the jet spreading rate, $K_q$ has been identified as the parameter of interest. The mixing of a turbulent jet with the surrounding fluid can be characterized by $K_q$ within the momentum dominated region of the flow. Experimental studies that have shown a range of measured round jet $K_{u_{y}}$ values of $\approx (0.052, 0.116)$, for a variety of density ratios, $s = \rho_{\mathrm{jet}}/\rho_{\infty}$, of $(0.14, 4.2)$~\cite{PanchapakesanLumley1993, Djeridane1996, Wygnanski1969, Chassaing1994} (See Fig.~\ref{fig:CL_SRvsFW}). These results imply a weak dependence on $s$, where, as $s$ increases, $K_{u_{y}}$ decreases \cite{Elkaroui2020}, although there is considerable variation in the experimentally measured values. Historically, spreading rates have also been calculated from simulations of turbulent jets. Many of these simulations solve the Reynolds Averaged Navier-Stokes (RANS) equations and use various turbulence models, the treatment of which has been shown to create systematic offsets in $K_q$ \cite{Magi2001, Pope1978}. LES's of round jets are less common. Although Wang et al. \cite{Wang2008} did not state spreading rates, the growth in the radial profiles of their He jet simulation ($s = 0.14$) was larger then that of the experimental values, with the $\mathrm{CO}_2$ ($s = 1.52$) and air ($s = 1.0$) jets being closer to the experiment. The calculations of Foysi et al. \cite{Foysi2010} for plane and round jets showed that for all $s$, the round jets had a $K_{u_{y}} = 0.116$, and plane jets had a $K_{u_{y}} = 0.112$, showing no dependence of $K_{u_{y}}$ on $s$. Although these studies do not state uncertainties or use validation methodologies in determining a level of agreement with experimental data, many state good agreement in downstream center-line quantities. This implies that jet spreading rates of variable density jets are a difficult parameter to reproduce computationally.

%%%%
% xRage
In this study, ILES simulations of the variable density turbulent \sfsix~jet of Charonko and Prestridge \cite{Charonko2017} will be computed with the multi-physics, compressible, hydrodynamic code, \xRage~\cite{Gittings2008}. The simulation will be compared to the experimental data with a focus on the turbulent mixing and jet structure. A validation study, as outlined by Wilson and Koskelo \cite{Wilson2020}, will be performed on the model with the parameters of interest being the jet spreading rates. 

This paper is structured as follows. In Sec.~\ref{sec:Meth}, the methods used in this study are described, including a description of the experimental data, the inflow methods used in the simulations, and the jet scaling and validation methodology. The results are presented in Sec.~\ref{sec:Res} where jet scalings are shown, the spreading rates are measured, and the validation Model Accuracy and Model Acceptance are examined. The behavior of the mixing in terms of the radial profiles of the higher-order statistics is also examined and compared to the experiment. Finally, the results are discussed in Sec.~\ref{sec:Disc}, where the Validation Evaluation and Validation Recommendations are presented.

%%%%%%%%%%%%%%%%%%%%%%%%%%%%%%%%%%%%%%%%%%%%
\section{Methods}
\label{sec:Meth}
%%%%%%%%%%%%%%%%%%%%%%%%%%%%%%%%%%%%%%%%%%%%

%%%%%%%%%%%%%%%%%%%%%%%%%%%%%%%%%%%%%%%%%%%%
\subsection{Turbulent Mixing Tunnel Experiment}
\label{sec:TMT}

% The apperatus
The turbulent mixing tunnel (TMT) experiment was designed to study low Mach number (Ma), variable density, turbulent mixing for statistically stationary turbulent flows \cite{Charonko2017, Gerashchenko2015, Lai2018}. The apparatus for the experiment consists of a downward-oriented jet within a wind tunnel. The flow of the wind tunnel (jet coflow) is in the direction of the jet so that mixed material is carried away from the observation locations. This allowed the experiment to be run for long periods of time to generate independent, instantaneous realizations of velocity and density fields for calculating the means and fluctuating turbulence quantities.

\begin{figure}[ht!]
\centering
\includegraphics[scale=0.3]{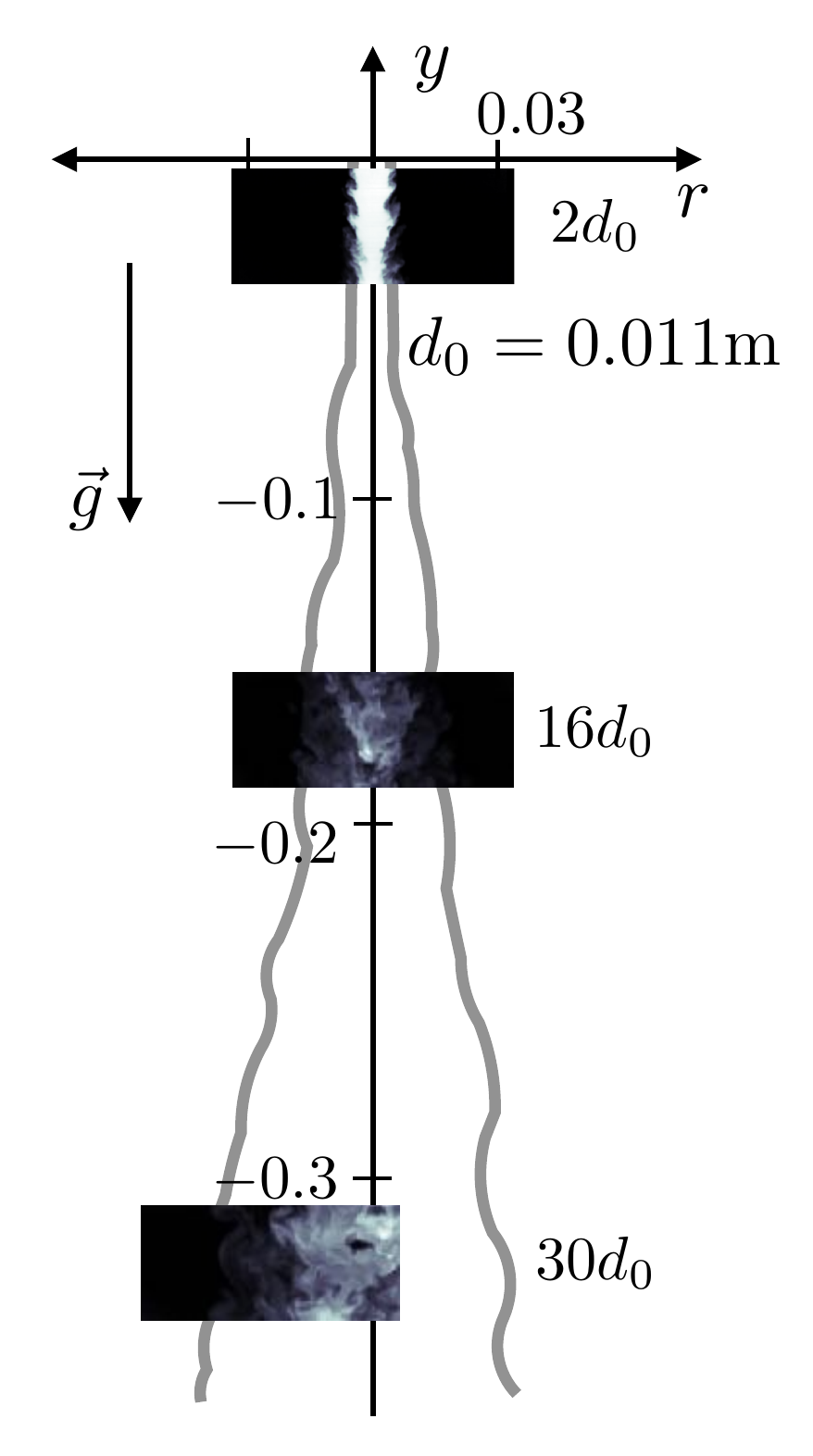}
\caption{A diagram representing the experimental data at the three downstream locations (\AD, \BD~and \CD~where $d_0 = 0.011 \mathrm{m}$ is the nozzle diameter). At each location, 10,000 simultaneous density and velocity fields are measured from the turbulent \sfsix~jet. Density fields are plotted here. On this diagram, $g$ is the local acceleration due to gravity and points downward in the direction of the \sfsix~jet.}
\label{fig:TMT_jet}
\end{figure}

% Charonko paper - The data
Charonko and Prestridge \cite{Charonko2017} studied the statistical properties of variable density turbulent jets by comparing a jet of \sfsix with coflowing air against an air jet with air coflow. At three downstream locations from the jet nozzle, 10,000 planar velocity and density fields were collected using particle image velocimetry (PIV) and planar laser-induced fluorescence (PLIF) (Fig.~\ref{fig:TMT_jet}). The three downstream locations are within the momentum dominated region of the jets and are taken at \AD~($y = (-0.005 \mathrm{m}, -0.035\mathrm{m})$), \BD~($y = (-0.16 \mathrm{m}, -0.19\mathrm{m})$), and \CD~$y = (-0.315 \mathrm{m}, -0.345\mathrm{m})$, where $d_{0} = 0.011 \mathrm{m}$ is the nozzle diameter. The \sfsix~jet gas consisted of a mixture of \sfsix, tracer particles, and, acetone vapor, and had a mixture density of $\rho_{0} = 3.9 \mathrm{kg/m}^3$ at the jet nozzle. This produced an Atwood number of $At = 0.62$ between the nozzle exit and the coflowing air (Tab.~\ref{tab:JetNums}). The air jet consisted of air, tracer particles, and acetone vapor, and in contrast, had $At = 0.09$. The two jets were matched on the bulk Reynolds number at the jet nozzles with $\mathrm{Re} \approx 20,000$, and $\mathrm{Ma} \approx 10^{-3}$. Particle fields and concentration images were taken from the flow approximately once per second, sampling the flow at a time scale larger than the integral time scale ($\approx 0.05 \mathrm{s}$), creating samples that are statistically independent in time. The dynamical properties of the variable density turbulence was then studied statistically. 

% Why is it good for validation
The data set produced by this experiment is a very large sample count, high spatial resolution ($dx = 4.865 \eta$, where $\eta$ is the Kolmogorov length scale measured at the nozzle, Tab.~\ref{tab:Case_data}) data set at a number of different downstream locations, with uncertainties for each density and velocity field. The data set is of subsonic, statistically stationary, turbulent mixing with a large density contrast and is well suited for validation efforts attempting to model such flows.

% Experimental Apparatus Diagram

% Dimensionless numbers for jet data
\begin{table*}[t]
\caption{Parameters of the \sfsix~jet. This data can also be found in Charonko and Prestridge \cite{Charonko2017}.}
\centering
\begin{tabular}{ l  c   c  c }
\hline
Mean jet exit velocity  &  $<u_{\mathrm{jet}}>$  &  6.60  &  $\mathrm{m/s}$  \\

Mean tunnel coflow velocity  &  $<u_{\mathrm{coflow}}>$  &  0.69  &  $\mathrm{m/s}$  \\ 

Jet mixture density  &  $\rho_{\mathrm{jet}}$  &  3.9  &  $\mathrm{kg/m}^3$  \\

Coflow air density  &  $\rho_{\mathrm{coflow}}$  &  0.92  &  $\mathrm{kg/m}^3$  \\ 

Jet Reynolds number, mean exit velocity  & $\mathrm{Re}_{\mathrm{jet}}$ & 19400 &  \\ [0.5ex] 

Jet Reynolds number, excess jet velocity  & $\mathrm{Re}_{\Delta u}$  & 17400 &   \\ [0.5ex] 

Taylor Reynolds number  & $\mathrm{Re}_{\lambda}$  & 62 &  \\ [0.5ex] 

Atwood number  & At  & 0.62 &   \\ [0.5ex] 

Mach number   & Ma & $10^{-3}$ &  \\ 

Kolmogorov microscale  &  $\eta$  &  59  &  $\mu \mathrm{m}$ \\
\hline
\end{tabular}
\label{tab:JetNums}
\end{table*}

%%%%%%%%%%%%%%%%%%%%%%%%%%%%%%%%%%%%%%%%%%%%
\subsection{\xRage~Simulation}
\label{sec:RAGE}

%%%%%%%%%%%%%%%%%%%%%%%%%%%%%%%%%%%%%%%%%%%%
\subsubsection{$\mathrm{\textit{SF}}_{6}$~Jet Simulations}
\label{sec:JetSimulations}

Two simulations of the \sfsix~jet were computed with the multi-material finite volume radiation-hydrodynamics code \xRage~\cite{Gittings2008}, one in 2D and the other in 3D. Table~\ref{tab:Case_data} describes the \xRage~simulations used in this study.

The simulations were performed by solving the Euler equations for compressible, inviscid flow on a rectangular Cartesian grid extending downward in the direction of the jet. At the top of the domain, the inflow was specified using an inflow model which we describe in Sec.~\ref{sec:inflow_model}. For the outflow boundary condition, we used a Dirichlet condition for the pressure, a piecewise constant extrapolation for the effective Gruenisen parameter \cite{Gruneisen1912} and a von Neumann condition (zero gradient) for all other primitive variables except for density, which was computed from the pressure, energy and extrapolated Gruenisen parameter \cite{Colonius2004}. At the time of writing, no perfect outflow boundary exists for compressible flow at low Ma. To minimize any influence of imperfect interactions between the boundary and the flow on the solution domain of interest, the mesh was gradually derefined to a downstream distance, so that the increased amount of numerical dissipation would result in close to laminar flow properties close to the outflow boundary. The sidewalls of the tunnel were modeled with a simple reflecting boundary condition. Initial tests showed wider jet spreading in the 2D simulation. Therefore, simulations in 2D employed a wider domain so that the jet flow would not hit the sidewalls within the downstream observation locations of the experiment. To again minimize any possible influence of flow-boundary interactions on the domain of interest, we used static mesh refinement to gradually de-refine the mesh close to the reflecting side wall boundaries.

% EOS gammas
The problem was simulated using a 2-material fluid model with a multi-gamma-law equation of state (EoS) and pressure-temperature equilibrium (PTE) mixed cell closure model \cite{Gittings2008}. The two fluids served as a representation of the jet and co-flow material, which are in reality each its own mixture of gases (Tab.~\ref{tab:gammas}). Therefore, the adiabatic constants, $\gamma$, for our two simulated fluids were estimated from the experimental data to represent the dynamics of the gas mixtures as closely as possible. The mole fractions of the constituent gases were found and the mixture heat capacities were calculated. The ambient conditions were taken as the average of the values of those recorded from the weather station within the laboratory on the days of experimentation. The average ambient conditions used to find the $\gamma$'s are $T=25.8^o \mathrm{C}$, $P = 79126 \mathrm{Pa}$ and $RH = 39.6\%$ (Relative Humidity).\footnote{The experiment was performed in Los Alamos NM in the summer. The ambient conditions within the laboratory fluctuated somewhat due to extreme weather changes as this is the rainy season and sudden rainstorms are common. Relative Humidity is most strongly affected and leads to an uncertainty in the Tunnel gas gamma of $0.14\%$ } 

% numerical methods
% sj
The Euler equations are integrated by computing fluxes at all cell faces using a piecewise parabolic spatial reconstruction of primitive variables and an HLLC Riemann solver. The fluxes in all dimensions are applied simultaneously and the system is integrated in time using a second order Runge-Kutta method.

\begin{table*}
\centering
\caption{Data from the experimental \sfsix~jet and the \simDD~and \simDDD~simulations. Here, $dx$ is the grid resolution of the data, $l_{\mathrm{grid}}$ is the grid cell resolution relative to the experimental data where $\eta = 0.059 \mathrm{mm}$, $T_{L_{11}}$ is the maximum integral time scale for the $16d_0$ and $30d_0$ locations, $T_{\mathrm{int}}$ is the number of integral time steps extracted from the data set, $(X,Z,Y)_{\mathrm{eff}}$ is the domain size after the inflow and outflow models are removed for the simulation, and the tunnel dimensions for the experiment, and $(X,Z,Y)$ is the domain size in grid cells of the simulations including the inflow and outflow models.}

\begin{tabular}{lccc}
% & & & \\
\hline
 & TMT~\sfsix & \xRage~\simDD & \xRage~\simDDD \\
\hline
ID & EXP. & \simDD & \simDDD \\
\hline
Dimension & 3D & 2D & 3D \\
$dx \:\: [\mathrm{mm}]$ & 0.28704 & 1.0352 & 1.3802 \\
$l_{\mathrm{grid}}$ & $1 \: (4.865 \eta)$ & 3.6063 (17.54 $\eta$) & 4.8084 (23.39 $\eta$) \\
$T_{L_{11}}$ & $0.05 \mathrm{s}$ & $0.07 \mathrm{s}$ & $0.05 \mathrm{s}$ \\
$T_{\mathrm{int}}$ & 10,000 & 103 & 60 \\
$(X,Z,Y)_{\mathrm{eff}} [\mathrm{m}]$ & 0.5, 0.5, 5.0 & 0.53, 0.50 & 0.18, 0.18, 0.52 \\
$(X,Z,Y) [\mathrm{gc}]$ & - & 512, 768 & 128, 128, 576\\
\hline
\end{tabular}
\label{tab:Case_data}
\end{table*}

\begin{table}
\centering
\caption{Adiabatic gas constants, $\gamma$, for the gas mixtures used in the \xRage~simulations.}
 \begin{tabular}{ c  c  c } 
 \hline
 Type & Gases &  $\gamma$ \\ [0.5ex] 
 \hline
 Air jet & Air, Acetone & 1.305 \\
 \sfsix~jet & \sfsix, Acetone & 1.100  \\
 Tunnel air & Air, $\mathrm{H}_{2} \mathrm{O}$ & 1.400  \\ 
  \hline
\end{tabular}
\label{tab:gammas}
\end{table}

% Filtered standard deviation fit parameters
\begin{center}
\begin{table}[ht]
\caption{Slope and y-intercept for the linear fits of standard deviation vs filter width for the coflow and jet core fluctuation PDFs (Fig.~\ref{fig:turb_model_sdvsfw}). Here $m$ is the slope of the line and $\sigma_{o}$ is the y-intercept.}
\centering
\begin{tabular}{ c | c  c }  
\hline
    &    $m \:\: [\mathrm{1}/\mathrm{s}]$    &    $\sigma_{o}\:\: [\mathrm{m}/\mathrm{s}]$   \\ [0.5ex] 
\hline
    & \multicolumn{2}{c}{coflow} \\

$u_{y, \mathrm{coflow}}^{\prime}$ & $-0.244$ & $0.0495$  \\  [1ex] 

$u_{r, \mathrm{coflow}}^{\prime}$ & $-0.313$ & $0.0595$  \\ 
\hline
    & \multicolumn{2}{c}{jet core} \\

$u_{y, \mathrm{jet}}^{\prime}$   & $-19.9$ & $0.415$  \\  [1ex] 

$u_{r, \mathrm{jet}}^{\prime}$ & $-17.4$ & $0.260$  \\ 
\hline
\end{tabular}
\label{tab:sigma_vs_filter_width_fits}
\end{table}
\end{center}

%%%%%%%%%%%%%%%%%%%%%%%%%%%%%%%%%%%%%%%%%%%%
\subsubsection{Inflow model}
\label{sec:inflow_model}

% Inflow model
Care must be taken in a simulation to specify the turbulent boundary conditions that best mimic the experimental conditions. The method used here is to implement a multi-scale, data-driven approach informed by the experimental data. Rather than trying to simulate the jet nozzle (as in Wang et al. \cite{Wang2008}), experimental data at the furthest upstream location can be modeled and used as an inflow boundary condition for the simulation. The model consists of filtering the experimental velocity to generate fluctuations on larger grid scales, then adding these fluctuations to the mean profiles. This method eliminates some of the complications in dealing with the near field jet nozzle flow \cite{Wang2008, Stanley2000, Bogey2010, Bres2018}, while still modeling the flow found in the TMT experiment.

\begin{figure}[ht!]
\centering
\includegraphics[width=0.95\linewidth,]{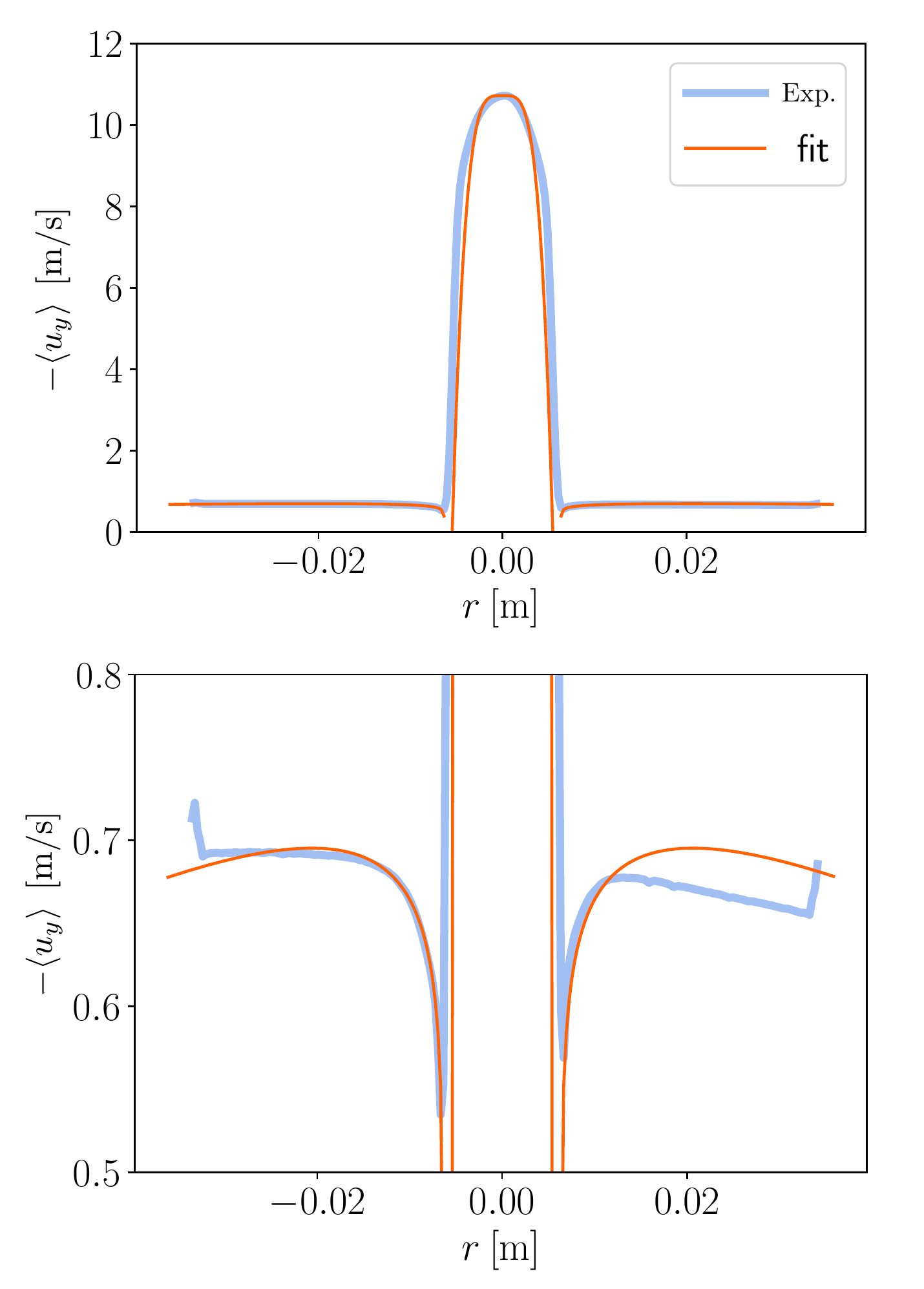}
\caption[]{The mean radial profile of the downstream velocity, $u_{y}$, across the jet at the closest position to the nozzle from the experimental data (blue). The fit to this data is given in Eq.~\ref{eq:sf6_fit} and was used as the mean component of the inflow for the simulation inflow model. Pure jet gas flows in through the central jet profile, and pure coflow gas flows in through the coflow regions on either side.}
\label{fig:turb_model_means}
\end{figure}

The furthest upstream location where data from the experiment was collected is at $0.005 \mathrm{m} (0.45 \mathrm{d}_{0})$ down from the nozzle. At this location, the jet and coflow have not significantly mixed, so the flow can be separated into jet flow (pure jet gas) and coflow (coflow air) regions (Tab.~\ref{tab:gammas}). The mean downstream velocity profiles near the jet exit are documented in Charonko and Prestridge \cite{Charonko2017} and are modeled well with a scaled \textit{Log-Law} profile. Because of a discontinuity at the centerline \textit{Log-Law} profile, we use a polynomial fit to the mean downstream velocity. These fits are presented in Eq.~\ref{eq:sf6_fit} and Fig.~\ref{fig:turb_model_means}, and are a model inflow for this validation study, not a generalized jet inflow model.

% SF6 mean inflow
\begin{table*}
\begin{equation}
\overline{u}_{\mathrm{y, BC}} = \begin{cases} 
-0.05 \ln(-r - 0.00635) - 3.5r - 0.98 & r < -0.00635 \mathrm{m} \\
0 &-0.00635 \mathrm{m} \leq r \leq  -0.0055 \mathrm{m} \\
(1.2 \times 10^{10}) r^{4} - 10.72 & -0.0055 \mathrm{m} < r < 0.0055 \mathrm{m} \\
0 & 0.0055 \mathrm{m} \leq r \leq 0.00635 \mathrm{m} \\
-0.05 \ln(r - 0.00635) + 3.5r - 0.98 & r > 0.00635 \mathrm{m}
\end{cases}
\label{eq:sf6_fit}
\end{equation}
\end{table*}

\begin{figure}[ht!]
\centering
\includegraphics[width=0.95\linewidth,]{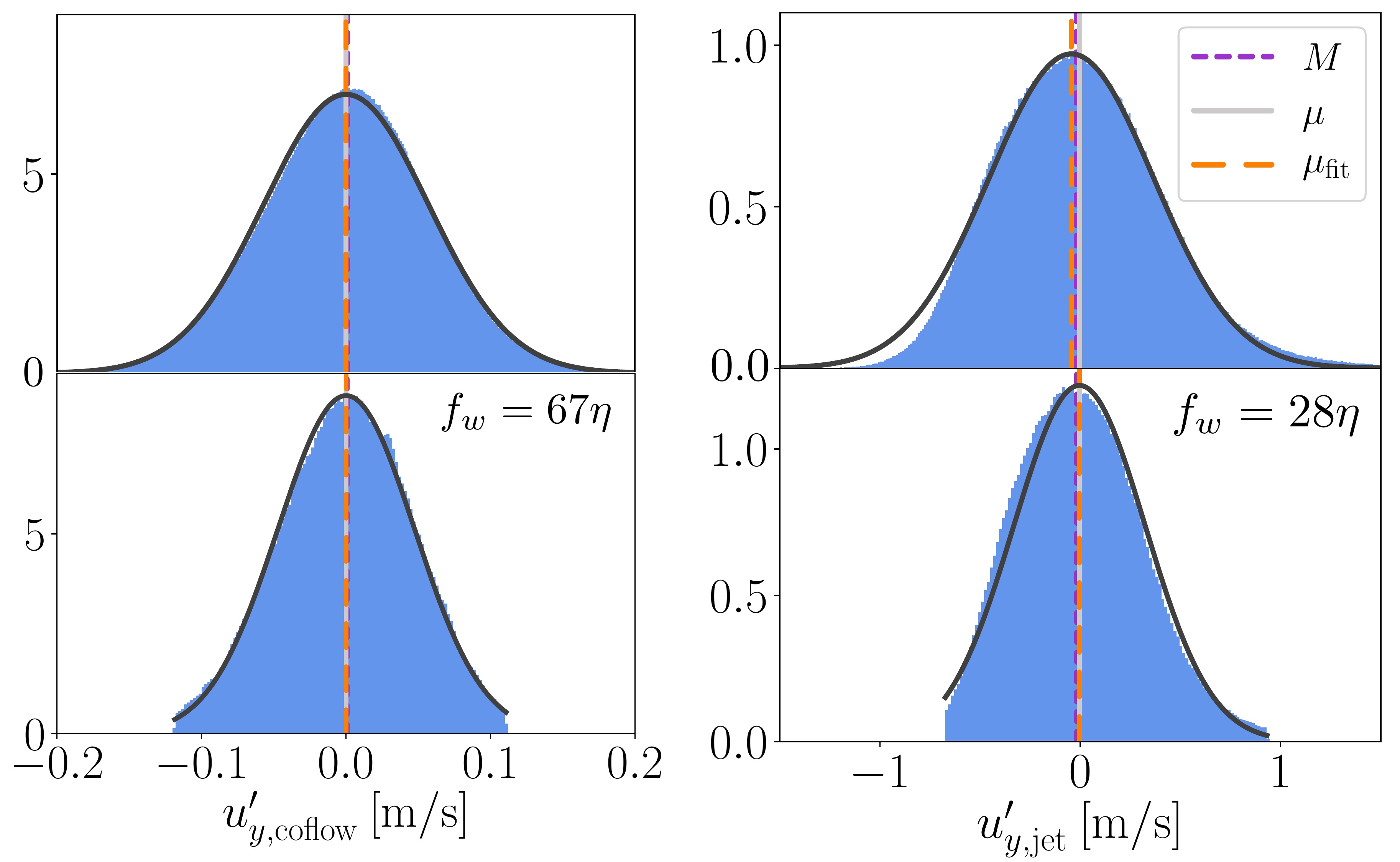}
\caption[]{The PDFs of downstream velocity fluctuations, $u^{\prime}_{y}$ in the coflow (left column) and jet core (right column) for the unfiltered (top row) and filtered (bottom row) experimental data. The black curve is a Gaussian fit with mean $\mu_{\mathrm{fit}}$, $\mu$ is the mean, and $M$ is the median of the data.}
\label{fig:turb_model_PDFs}
\end{figure}

\begin{figure}[ht!]
\centering
\includegraphics[width=0.95\linewidth,]{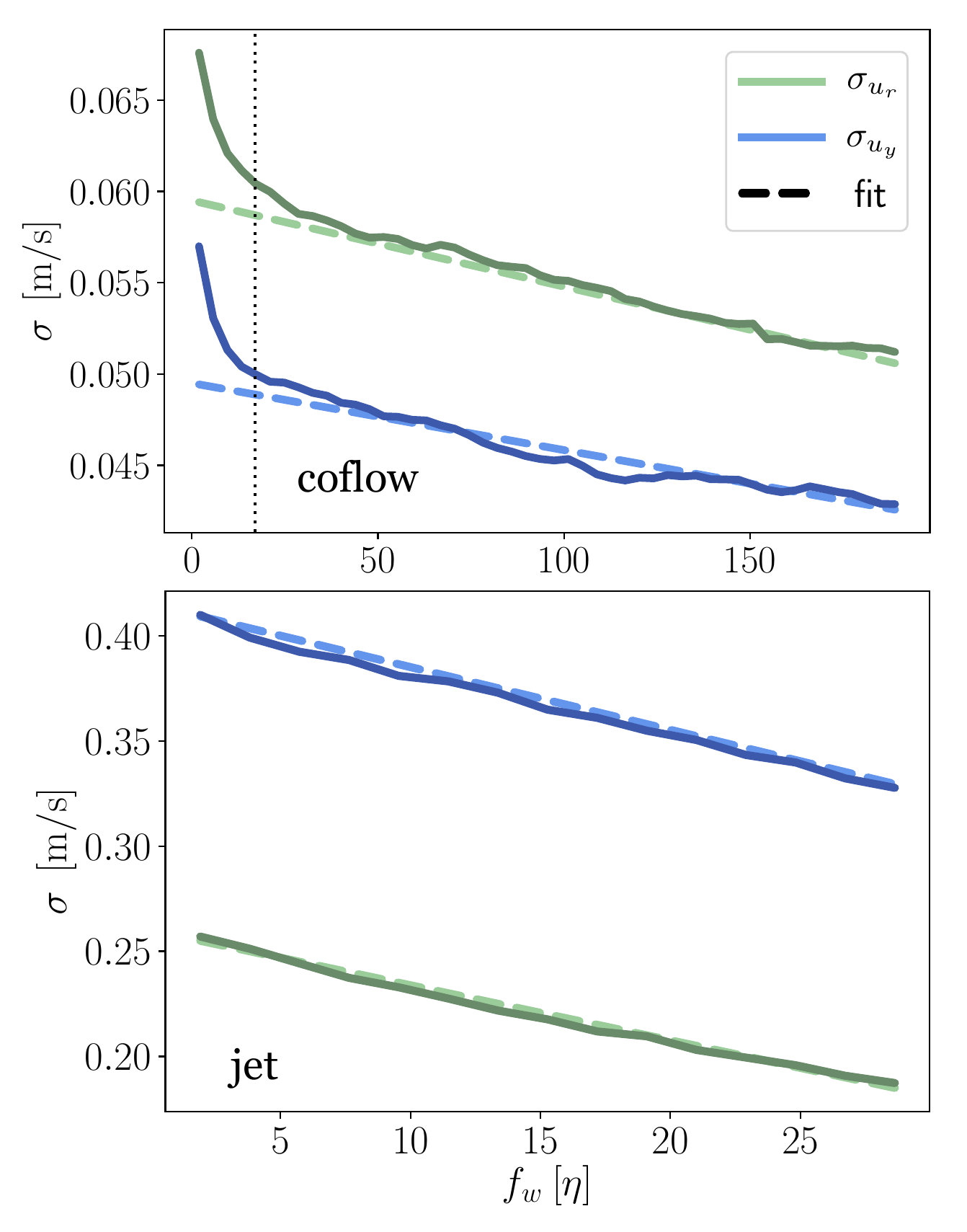}
\caption[]{Standard deviation $\sigma$, vs filter width, $f_{w}$, in units of Kolmogorov length scales from the experiment, used for the inflow model. The black dotted line in the upper panel shows the grid spacing of the highest spatial resolution simulation ($17.5 \eta$ for \simDD)}
\label{fig:turb_model_sdvsfw}
\end{figure}

To generate the fluctuations, probability distributions functions (PDFs) for the coflow ($r \approx (-0.03\mathrm{m}, -0.015\mathrm{m}), (0.015\mathrm{m}, 0.03\mathrm{m})$ where $0.03\mathrm{m} \approx 2.7\mathrm{d}_{0}$ and $0.015\mathrm{m} \approx 1.4\mathrm{d}_{0}$) and jet core ($r \approx (-0.005\mathrm{m}, 0.005\mathrm{m})$ where $0.005\mathrm{m} \approx 0.45 \mathrm{d}_{0}$) velocities were found and filtered to various length scales. Figure~\ref{fig:turb_model_PDFs} gives the PDFs for the axial velocity fluctuations ($u_{y}^{\prime}$) over the coflow (left column), over the jet at the boundary (right column), for data on the experimental grid (top row), and at selected filtered length scales (bottom row). A 1D, cross-stream filter with a Gaussian filter kernel was used to produce this length scale decomposition \cite{Getreuer2013}. After filtering, the distributions remain fairly Gaussian for the most probable fluctuations, although the characteristics skew of turbulent PDFs in a mean shearing velocity gradient become more apparent \cite[p.~173]{LaRue1974, Wygnanski1969, Pope2000}. The widths of the Gaussian fits of these distributions can be characterized by the standard deviation. Figure~\ref{fig:turb_model_sdvsfw} shows the standard deviation of the PDF fits of the filtered fluctuations with respect to filter width for both the jet and coflow. The standard deviations of the filtered PDFs from the jet core are linear from the grid-scale up to $\approx 30 \eta$. In the coflow, for filter widths above $\approx 17 \eta$ (upper panel, vertical dotted line), $\sigma(f_{w})$ is very nearly linear. For the smallest filter widths, below $17 \eta$, $\sigma(f_{w})$ is not linear and obtains the maximum value at the grid-scale \footnote{This is due to how the PIV algorithm interacts with under-seeded regions of the flow. In the coflow, tracer particles and the background air are mixed by the turbulence in the wind tunnel far upstream of the jet nozzle. This mixing is imperfect and small regions of the coflow without tracer particles persist to the observation regions. On the edges of these under seeded regions, the velocities given by the PIV algorithm can be larger than expected, due to groups of tracer particles not being tracked properly. Some of this data can be removed in post-processing steps, but values that can't be easily identified as outliers remain. Because these velocities are generally found at or near the grid-scale, they are removed from the data by filtering at the low filter widths.}. Simulations were not calculated for grid resolutions in the non-linear region.

The linear fits to the $\sigma$ vs. filter width profiles in Fig.~\ref{fig:turb_model_sdvsfw} are used to specify the velocity fluctuation for the inflow model. The parameters for these linear fits are found in Tab.~\ref{tab:sigma_vs_filter_width_fits}. For a given grid cell size on the simulation boundary, $dx$, a Gaussian PDF defined by $\sigma(dx)$ is generated by the linear fits. This PDF is then randomly sampled every time step to generate the velocity fluctuation, $u^{\prime}_{y, i}$. Then the velocity at that grid cell is taken as the mean velocity in that region given by Eq.~\ref{eq:sf6_fit}, plus the fluctuation, $u = \overline{u}_{y, \mathrm{BC}} + u^{\prime}_{y, i}$. Time and space correlations in the velocity fluctuations from the experimental data set were not considered. For the purposes of validation against experimental data, this data driven approach to modeling the turbulent inflow eliminates some of the complications in simulating the dynamics near the jet nozzle.

%%%%%%%%%%%%%%%%%%%%%%%%%%%%%%%%%%%%%%%%%%%%
\subsection{Scaling and Validation methods}
\label{sec:val}

%%%%%%%%%%%%%%%%%%%%%%%%%%%%%%%%%%%%%%%%%%%%
\subsubsection{Half Width Virtual Origin Scaling}
\label{sec:VOSR}

% Same DS data - Fr scaling, VO SR method
The inflow model presented in Sec.~\ref{sec:inflow_model} simulates the jet inflow without modeling a physical jet nozzle. When comparing the simulated jet data with the experimental data, a common comparison would use the downstream position scaled by nozzle diameter \cite{Wang2008, Elkaroui2020}. Because this study is focused on the downstream variable density turbulent mixing between the jet and the coflow (at the \BD~and \CD~locations), a comparison method that focuses on the dynamics of the downstream mixing is more appropriate. The scaling method used in this work shifts the data using the virtual origins, $y_{q, 1/2}$, determined by the downstream behavior of the flow variables $q = \rho$ or $u_{y}$. The Spreading rate, $K_{q}$, the jet half-width, $r_{q, 1/2}$, and $y_{q, 1/2}$ are linearly related in the momentum dominated regime and are related by (Eq. 3.5 of \cite{Charonko2017}) \cite[Ch 5.2.3]{Wygnanski1969, Chassaing1994, Djeridane1996, Pope2000}

\begin{equation}
r_{q, 1/2} = K_{q} (y - y_{q, 1/2})
\label{eq:VOSR}
\end{equation} 

\noindent
where $q$ is one of the flow variables characterizing jet behavior (typically $u_{y}$ or Mass Fraction). By determining values for $K_{q}$ and $r_{q, 1/2}$ from within the downstream shearing region, a value for $y_{q, 1/2}$ can be found and used to compare the experimental and simulated data sets. This method focuses the comparison on the jet spreading behavior of particular downstream locations and decreases the dependence on the upstream jet flow details. It also uses the flow properties in the shearing region for the scaling, rather than the centerline values. Other virtual origin definitions exist that rely on the centerline velocity decay \cite{Chen1980, Charonko2017}, which can be difficult to use when dealing with under-sampled data. This shear region scaling method is useful when dealing with under-sampled 3D data as rotating or slicing the domain can increase the effective sample count further away from the centerline.

%%%%%%%%%%%%%%%%%%%%%%%%%%%%%%%%%%%%%%%%%%%%
\subsubsection{Validation Methodology}
\label{sec:uct}

The simulations will be investigated following the validation methodology of Wilson and Koskelo \cite{Wilson2020}, customized for this statistically stationary turbulent jet study. The quantities of interest used to assess the Model Accuracy are the spreading rates, $K_q$, as defined by the jet half-width. In general, validation metrics are not limited to single sets of parameters and can include multiple parameters of interest \cite{Maupin2018}. To limit the scope of this study, and to focus the diagnostics on variable density mixing, the spreading rates are the only parameters of interest evaluated with the validation metric. The predictive accuracy of $K_q$ is defined as

\begin{equation}
\delta_{K_q} \geq e_{K_q} \pm \Delta_{K_q}
\label{eq:PA}
\end{equation} 

\noindent
where $e_{K_q}$ is validation comparison error, and $\Delta_{K_q}$ is the validation uncertainty for the parameter of interest $K_q$ \cite{ASME2019}. The validation comparison error is given in terms of a validation metric, in this case, the absolute difference,

\begin{equation}
e_{K_q} = d_{1}(K_{q}) = | K_{q, \mathrm{Exp}} - K_{q, \mathrm{Sim}} |.
\label{eq:D_one}
\end{equation}

The validation uncertainty, $\Delta_{K_q}$, quantifies the uncertainty on $K_q$ through out the validation process and consists of four main components, the experimental uncertainty, $\delta_{\mathrm{Exp}}$, the numerical uncertainty, $\delta_{\mathrm{Num}}$, the uncertainty in the simulation input parameters, including the initial and boundary conditions, $\delta_{\mathrm{IC}}$, and the uncertainty in the analysis method used to compare the simulation and experimental data, $\delta_{\mathrm{Comp}}$ \cite{Wilson2018}. If the uncertainties are uncorrelated, $\Delta_{K_q}$ is the square root of the sum of squares of these uncertainty components \cite{ASME2019}.

The interpretation of validation metrics in terms of the input data can be difficult as quantities from different sources are generally combined into a limited set of values. Examination of the ratio of the validation comparison error and the validation uncertainty has been shown to give some bounds on the meaning of the predictive accuracy \cite{Wilson2018}. If $e_{K_q}/\Delta_{K_q} \leq 1$, the validation uncertainty is too large and the predictive accuracy is ambiguous. If $e_{K_q}/\Delta_{K_q} >> 1$, the predictive accuracy indicates a systematic error, such that the simulated results are fundamentally different from the experiment.

%%%%%%%%%%%%%%%%%%%%%%%%%%%%%%%%%%%%%%%%%%%%
\section{Results}
\label{sec:Res}
%%%%%%%%%%%%%%%%%%%%%%%%%%%%%%%%%%%%%%%%%%%%

This section presents the comparison between the two simulated \xRage~\sfsix~jets and the experimental \sfsix~jet from Charonko and Prestridge \cite{Charonko2017}. The simulations produced for this study are presented in Sec.~\ref{sec:Simulations}. The half-width virtual origin scaling is applied in Sec.~\ref{sec:DSScaling} to find the comparable kinematic regions between the jets. In Sec.~\ref{sec:SR} the Model Accuracy and Model Acceptability for the jet spreading rates are tested (Sec.~\ref{sec:SR:Val}) and the simulated spreading rates are compared to those in the literature (Sec.~\ref{sec:SR:lit}). In an attempt to understand the differences between the experimental and simulated spreading rates, Sec.~\ref{sec:jet_diag} examines the self similarity scaling (Sec.~\ref{sec:Self_Similarity_Scaling}) and the free shear layer statistics for the scaled radial profiles (Sec.~\ref{sec:Rprof_Stats}).

% The downstream centerline scaling is shown in Section~\ref{sec:CLScaling}.

%%%%%%%%%%%%%%%%%%%%%%%%%%%%%%%%%%%%%%%%%%%%
\subsection{Simulations}
\label{sec:Simulations}

\begin{figure}[ht!]
\centering
\includegraphics[width=0.95\linewidth,]{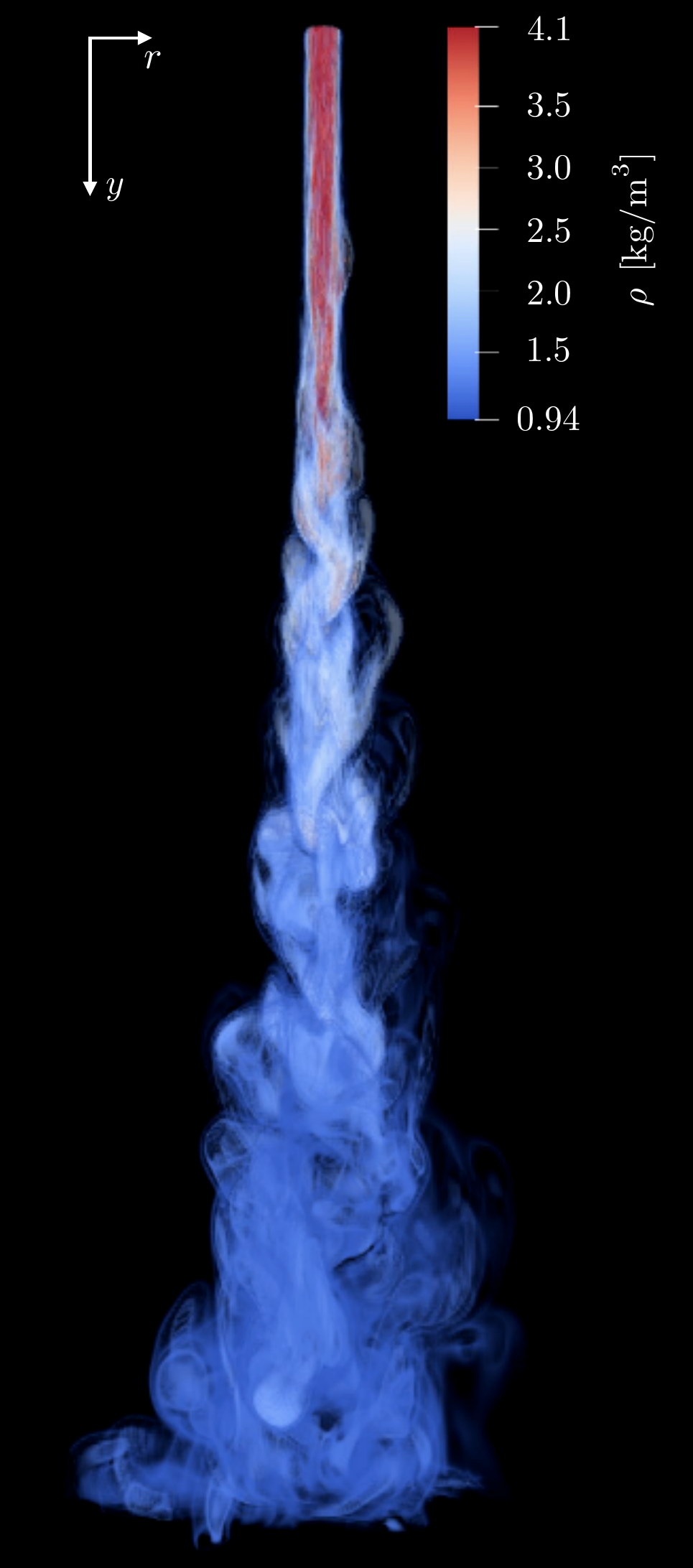}
\caption[]{Three dimensional density rendering of the \simDDD~simulation.}
\label{fig:072_density}
\end{figure}

% Domain size, dimension, number of time steps...
Two simulations were computed for this study, one in 2D and the other in 3D. Figure~\ref{fig:072_density} shows a 3D density rendering for one time step of the \simDDD~simulation. The inflow and outflow models of each simulation were removed from the top and bottom of the domain before any detailed comparison was made. The \simDDD~simulation was run with a domain of $128 \times 128 \times 576$ grid cells, producing usable data with physical dimensions of $0.18\mathrm{m} \times 0.18\mathrm{m} \times 0.52 \mathrm{m}$ ($16.4\mathrm{d}_{0} \times 16.4\mathrm{d}_{0} \times 47.3 \mathrm{d}_{0}$) (Tab.~\ref{tab:Case_data}). This simulation has a grid resolution of $dx_{3D} = 1.3802 \mathrm{mm}$ or $23.39 \eta$. The Kolmogorov length scale ($\eta = 0.059 \mathrm{mm}$) is measured from the experiment at the centerline near the jet nozzle, and is used as a common scaling between the simulations and experiment as it is the lower length scale limit for ILES. The \simDD~simulation was run with a domain of $512 \times 768$, with a usable physical dimension of $0.53 \mathrm{m} \times 0.50 \mathrm{m}$ ($48.2 \mathrm{d}_{0} \times 45.5 \mathrm{d}_{0}$) and $dx_{2D} = 1.0352 \mathrm{mm} \:\: (17.54\eta)$. The \simDD~simulation required a much wider domain than the \simDDD~simulation due to the increased jet spreading in two dimensions. The \simDD~simulation was run for $3 \times 10^6$ time steps and the \simDDD~simulations was run for $2 \times 10^6$ time steps. Once the initial transients were removed from the simulations by allowing the total kinetic energy to reach a steady state, the data was sampled at integral time steps ($\approx 0.05 \mathrm{s}$) generating 60 statistically independent time steps for the \simDDD~simulation and 103 for the \simDD~simulation. The number of grid cells and time steps of the \simDDD~simulation are similar to those of Wang et al. \cite{Wang2008} with $t_{\mathrm{aver}} = 1090$ (in their notation), where the LES's of Wang et al.\cite{Wang2008} were computed on a cylindrical grid. 

% Filtering
The data from the experiment is represented on a grid of $dx_{\mathrm{Exp}} = 0.28704 \mathrm{mm} \: (4.865\eta)$, much smaller than the simulations. In an attempt to attenuate implicit filtering effects, both the ILES data and the experimental data were filtered to a common grid scale. The filtering method used a 2D Gaussian filter kernel \cite{Getreuer2013}. The simulation data was filtered to 9 grid cells, and the experimental data was filtered to match these length scales. For the remainder of this paper, angle brackets ($\langle q \rangle$) denote Reynolds averages taken on the grid, and over-bar notation ($\overline{q}$) denotes filtered quantities.

\begin{figure}[ht!]
\centering
\includegraphics[width=0.95\linewidth,]{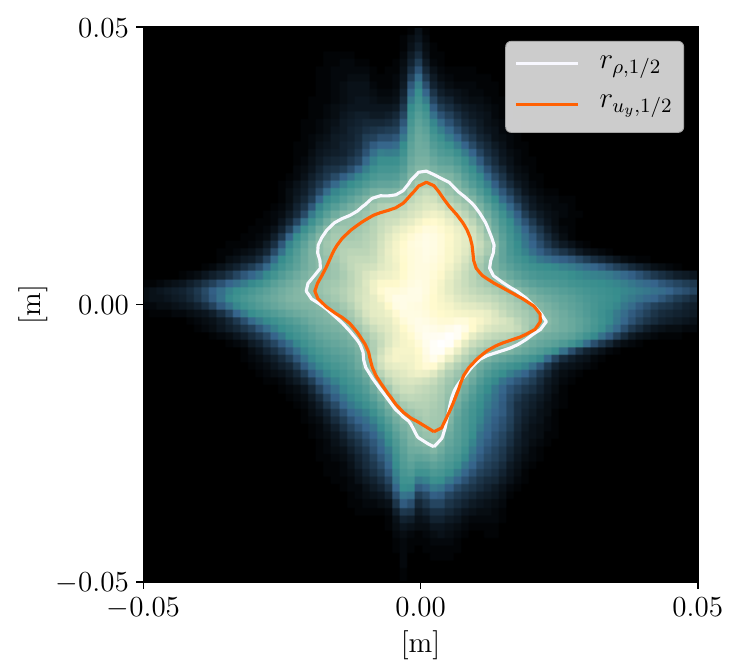}
\caption[]{Cross stream slice of the average density for the \simDDD~simulation with the density (white line) and downstream velocity (orange line) half-widths. The grid imprints can be seen aligned with the grid. The variation in the half-widths reflects the variation in the spreading rates calculated from different slicing angles.}
\label{fig:halfwidth_slice}
\end{figure}

% grid imprints on 3D sim
The computation of turbulent ILES's, in general, requires a high spatial resolution to reduce the influence of the grid geometry on the large scales of the flow. For statistically stationary turbulent flows, the sample count must be sufficiently large to approach stochastic convergence, at least in the mean quantities. Both of these conditions are required to statistically compare an ILES to a real fluid flow and produce meaningful, consistent results. Unfortunately, the \simDDD~simulation does not have sufficient spatial resolution to overcome the influence of the computational grid on the large scales (Fig.~\ref{fig:halfwidth_slice}). Grid artifacts are seen along the cross-stream grid axes in both density, $\rho$, and velocity, $u$. Ordinarily this would conclude a comparison study looking for evidence that the computed solution is representative of the physical flow, and the simulation would be run again with higher resolution. But the goal here is to improve jet comparison methodologies by putting them in a formal validation framework. A validation study on this under-resolved simulation is still useful as it outlines methodology for further studies of turbulent variable density jets. As we shall see, the validation workflow of Wilson and Koskelo \cite{Wilson2020} is able to provide evidence of these resolution issues, potentially useful for studies where the grid imprints are not so obvious. This under resolved \simDDD~simulation was the highest resolution simulation that could be afforded given the time sampling requirements, so a resolution study was not performed.

% Constructing the comparable data sets - 3D data sampling / slicing
The experimental data set consists of images of density and tracer particle fields, illuminated by laser sheets oriented vertically that sliced through the centerline of the jet. To replicate the form of the experimental \sfsix~jet data set as closely as possible, rather than computing azimuthal averages for the \simDDD~simulation, the volume was sliced at an angle, $\theta$ (azimuthally in cylindrical coordinates), creating a set of 2D axial samples down the length of the jet. For each of the 60 integral time steps, the volumes were slices at angles from $\theta = 0$ to $15\pi/16$ at azimuthal increments of $\Delta \theta = \pi/16$, creating 16 cross stream slices. 

The method of generating uncertainties by subsampling the simulated data in time \cite{Wilson2020} proved difficult because all of the integral time steps were required to produce smooth mean radial profiles for $\overline{\rho}$, $\overline{u}_{y}$ and $\overline{u}_{r}$. Using the standard deviation for the uncertainty in certain quantities also generated prohibitively large uncertainty ranges. This is due to the low sample count and high intermittency of the simulation data. The intermittency defined by the variance of a flow variable is also a characteristic of turbulence and requires its own comparison. Because these methods of generating uncertainty in the simulation data were problematic, rather than averaging these 16 slices, a representative sampling angle of $\theta = \pi/8$ was chosen for the comparison. The remaining slices were used to quantify the maximum and minimum variation from the $\theta = \pi/8$ sample in the off axis profiles. For radial profiles, these samples are then flipped across the centerline to form an average. If at the same time step a slice through the domain can be flipped to create two independent samples, then the effective sample count doubles, from 60 to 120. This assumption is not valid near the centerline but increases the effective sample count further into the shearing region. The \simDD~data was not given an uncertainty.

%%%%%%%%%%%%%%%%%%%%%%%%%%%%%%%%%%%%%%%%%%%%
\subsection{Jet Half-width Virtual Origin Scaling}
\label{sec:DSScaling}

% Section intro
In this section, the downstream virtual origin, $y_{u_{y}, 1/2}$, scaling from Sec.~\ref{sec:VOSR} will be used to find comparable kinematic regions between the simulations and the experimental data sets. The half-width profiles for the simulations are found for downstream regions of the jets, away from the nozzle, reducing the dependence of Eq.~\ref{eq:VOSR} on the inflow model, and focusing on the mixing within the shearing region.

\begin{figure}[ht!]
\centering
\includegraphics[width=0.95\linewidth,]{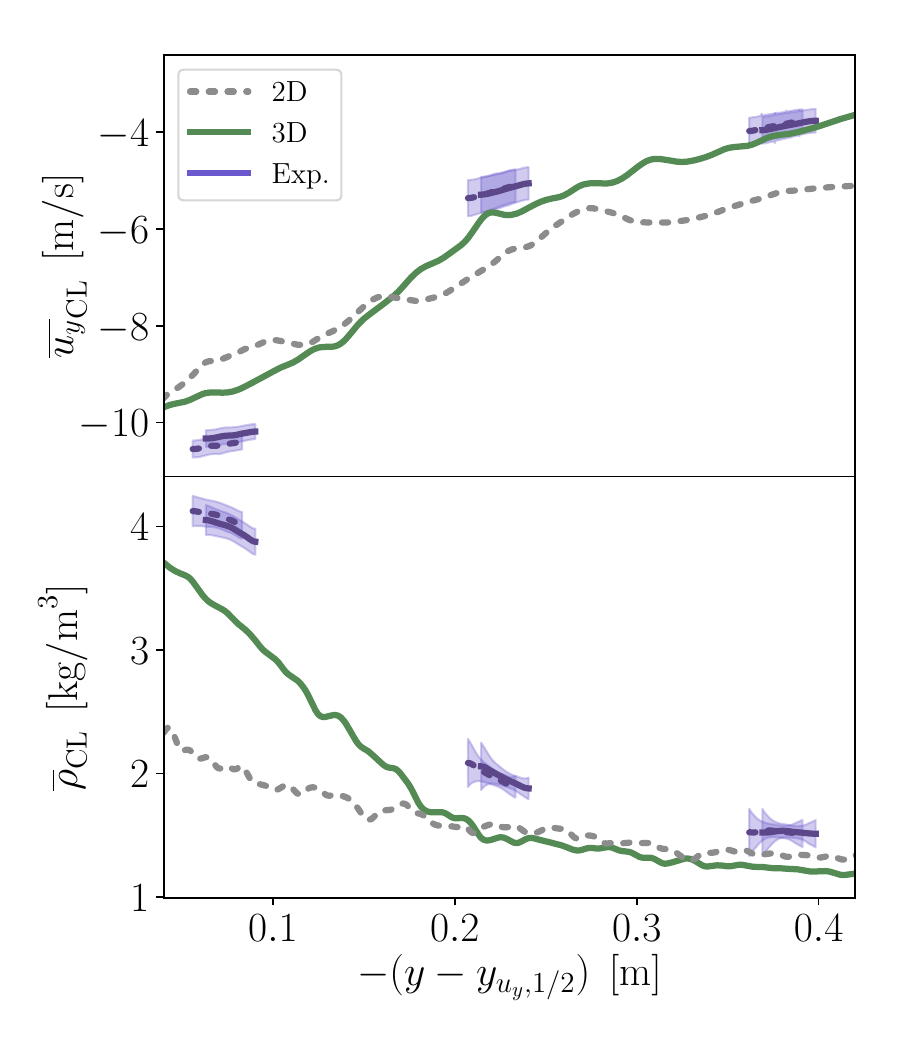}
\caption[]{Centerline Reynolds averaged downstream velocity (Top Panel) and density (Bottom Panel) for the experiment (purple), \simDD~simulation (grey) and the \simDDD~simulation (green). The purple shaded area is the uncertainty in the experimental data. The data has been filtered and aligned on velocity virtual origin, $y_{u_{y}, 1/2}$. The x-axis is the, $y_{u_y, 1/2}$, subtracted downstream distance.}
\label{fig:filtered_CL_profiles}
\end{figure}

Figure~\ref{fig:filtered_CL_profiles} shows the centerline profiles for $\overline{\rho}_{\mathrm{CL}}$ and $\overline{u}_{y, \mathrm{CL}}$. The centerline profiles have been filtered to similar grid spacing and scaled on downstream velocity virtual origin, $y_{u_{y}, 1/2}$. The uncertainties shown for the experimental regions are the measurement uncertainties of the fields at the centerline summed as the square root of the sum of squares over the image samples. The difference in the downstream shift in the experimental profiles with respect to the the \simDD~(dashed line) and \simDDD~(solid line) simulations are caused by $y_{u_{y}, 1/2}$ changing with respect to the two different matching filter widths for each simulations. The $y_{u_{y}, 1/2}$ increases with filter width, which causes comparable locations to be shifted downward in Fig.~\ref{fig:filtered_CL_profiles}. The \simDDD~simulation has $y_{u_{y}, 1/2} = -0.0615 \mathrm{m} (-5.591\mathrm{d}_{0})$ with a matching filtered experimental value of $y_{u_{y}, 1/2} = 0.0523 \mathrm{m} (4.755 \mathrm{d}_{0})$, and the \simDD~simulation has $y_{u_{y}, 1/2} = -0.1007 \mathrm{m} (9.155\mathrm{d}_{0})$ with a matching filtered experimental value of $y_{u_{y}, 1/2} = 0.0451 \mathrm{m} (4.1 \mathrm{d}_{0})$. 

Scaling the data with $y_{u_{y}, 1/2}$ acts to shift the $\overline{u}_{y}$ profile for the simulations and the experimental data at \Bd~and \Cd~closer together to better match the behavior in the mixing region, where \Ad, \Bd~and \Cd~are the half width scaled observation location labels of the TMT experiment (corresponding to \AD, \BD~and \CD~for the unshifted locations). In the \simDDD~simulation, at \Cd, the $\overline{u}_{y, \mathrm{CL}}$ is consistent with the experimental profile. At \Bd, for both the \simDDD~and \simDD~simulations, the fluid is still traveling downwards faster than in the experiment, although the \simDDD~simulation is close to the experiment, being $\approx 0.2 \mathrm{m/s}$ outside of the uncertainty range. None of the $\overline{\rho}_{\mathrm{CL}}$ profiles at all downstream locations for both \simDDD~and \simDD~fall within the uncertainty of the experimental data, although, at the \Cd~location, they're within $\approx 0.5 \mathrm{kg/m}^3$. In the \simDDD~simulation, this could partially be due to a persistent kink in the $\overline{\rho}_{\mathrm{CL}}$ profile ($-(y - y_{u_y, 1/2}) \approx 0.18 \mathrm{m}$ in Fig.~\ref{fig:filtered_CL_profiles}), most likely due to the simulation resolution. 

\begin{figure*}[ht!]
\centering
\includegraphics[width=0.95\linewidth,]{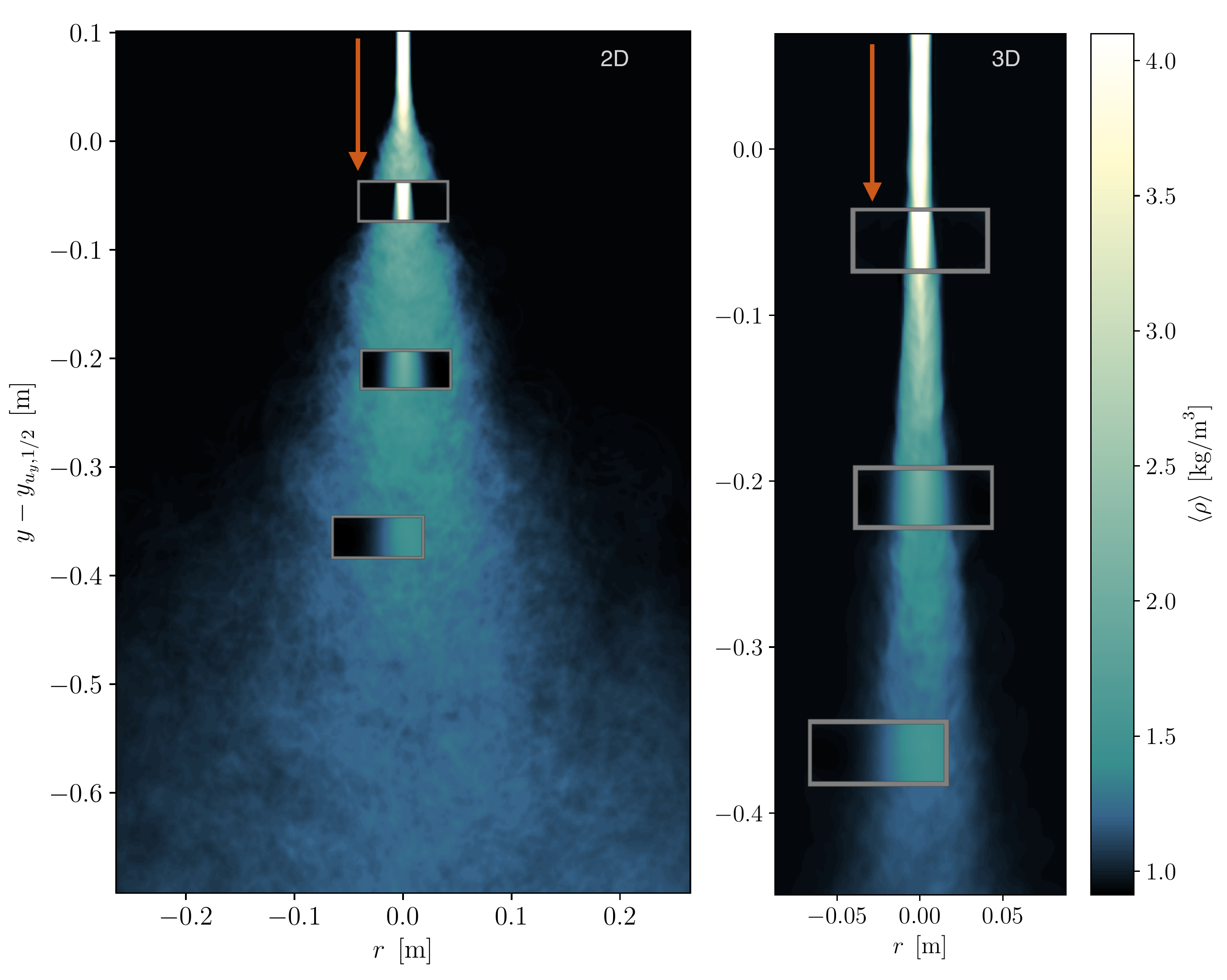}
\caption[]{Reynolds averaged density fields for the \simDD~simulation (left) and the \simDDD~simulation (right) with the experimental Reynolds averaged density, shown in boxes at the $y_{u_{y}, 1/2}$ scaled locations downstream of the jet nozzle.}
\label{fig:072_067_density}
\end{figure*}

 % R prof 2D
The $y_{u_{y}, 1/2}$ scaling is shown graphically for the density fields in Fig.~\ref{fig:072_067_density} for both simulations. The scaling does not produce a reasonable agreement in the density fields between the \simDD~simulation and the experiment due to the differences between plane and round jets. The $y_{q, 1/2}$ scaling checks the kinematic behavior of the jet spreading within the momentum-dominated regime. For the \simDDD~simulation, the $y_{u_{y}, 1/2}$ scaling ignores the long inviscid core caused by insufficient grid cell resolution, aligning the radial behavior of the simulation and experimental fields quite well.

%%%%%%%%%%%%%%%%%%%%%%%%%%%%%%%%%%%%%%%%%%%%
 \subsection{Jet Spreading Rates}
 \label{sec:SR}

\begin{table*}
\caption{Spreading rates, $K_q$, from the experiment and the simulations for $u_{y}$ and $\rho$, and values from the Validation Accuracy assessment (Sec.~\ref{sec:uct}). For the validation uncertainty, uncertainty in the \simDDD~simulation is taken as $2\sigma$ of the slicing sample distribution. Both \simDD~and \simDDD~are filtered to 9 grid cells, which are different sizes for each simulation. Corresponding experimental spreading rates are denoted by (2D grid, 3D grid), although the values are very similar. The percentages in brackets for the validation comparison error, $e_{K_q}$, are the percent differences between the experimental and simulation spreading rates.}
\small
\centering
\begin{tabular}{lccc}
\hline
  & Exp. & \simDD & \simDDD \\
\hline
 & (\simDD~grid, \simDDD~grid) & &  \\
\hline
$K_{u_y}$ & ($0.048 \pm 0.008$, $0.048 \pm 0.008$) & 0.203 & $0.064^{+0.010}_{-0.004}$ \\ 
 $K_{\rho}$ & ($0.055 \pm 0.004$, $0.054 \pm 0.004$) & 0.215 & $0.078^{+0.006}_{-0.000}$ \\
 $e_{K_{u_y}}$ & - & 0.155($310\%$) & 0.016(33\%) \\
 $e_{K_\rho}$ & - & 0.160(286\%) & 0.024(45\%) \\
 $\Delta_{K_{u_y}}$ & - & 0.011 & 0.023 \\
 $\Delta_{K_{\rho}}$ & - & 0.0073 &  0.018 \\
 $e_{K_{u_y}}/\Delta_{K_{u_y}}$ & - & 14.2 & 0.71 \\
 $e_{K_\rho}/\Delta_{K_\rho}$ & - & 21.9 & 1.34 \\

\hline
\end{tabular}
\label{tab:SR}
\end{table*}

In this section, the uncertainty in the spreading rate calculation is discussed (Sec.~\ref{sec:SR:uct}), and the validation Model Accuracy outlines in Sec.~\ref{sec:uct} will be applied to the spreading rates computed from the simulations. The simulated spreading rates will then be examined in terms of the values found in the literature (Sec.~\ref{sec:SR:lit}). Finally, the differences in behavior between the velocity and density spreading is investigated (Sec.~\ref{sec:SR:Be}).

%%%%%%%%%%%%%%%%%%%%%%%%%%%%%%%%%%%%%%%%%%%%
\subsubsection{Error propagation and uncertainty in the $K_q$ calculations}
\label{sec:SR:uct}

The spreading rate, $K_q$, is found by fitting Eq.~\ref{eq:VOSR} to the jet half-width defined by the flow variable, $q$ (Sec.~\ref{sec:VOSR}). Table~\ref{tab:SR} gives the $K_q$'s for $u_{y}$ and $\rho$ calculated from the experiment and the simulations. Least squares regression \cite{Taylor1997} was used to fit Eq.~\ref{eq:VOSR}, and produced r squared values $> 0.94$ in all cases. 

Uncertainty in the experimental spreading rates is calculated as

\begin{equation}
\delta K_q = \sqrt{(2\sigma_{\mathrm{lin fit}})^2 + \delta K_{q_{\mathrm{rel}}}^2}
\end{equation}

\noindent
where $\sigma_{\mathrm{lin fit}}$ is the standard error of least squares regression, and $\delta K_{q_{\mathrm{rel}}}$ is the uncertainty is $K_q$ from the $q$ field. $\delta K_{q_{\mathrm{rel}}}$ is calculated by propagating the relative uncertainty in $r_{q, 1/2}$ through Eq.~\ref{eq:VOSR}, where the relative uncertainty in $r_{q, 1/2}$ is taken as the uncertainty in the $q$ field at $r_{q, 1/2}$. 
%The non-linear correlation terms in the error propagation calculations were not used.

For the \simDDD~simulation, the uncertainty associated with the spreading rate was taken as the square root of the sum of squares of the sampling angle uncertainty and the standard error of least squares regression for the $\pi/8$ sample. The sampling angle uncertainty was taken as the difference between the $\pi/8$ sample and the maximum and minimum spreading rates calculated for each sampling angle. The sampling angle uncertainty produced asymmetric bounds on $K_q$ due to the grid imprints. This is problematic for validation metrics and will be discussed in the following section.

%%%%%%%%%%%%%%%%%%%%%%%%%%%%%%%%%%%%%%%%%%%%
\subsubsection{Validation Accuracy and Acceptability}
\label{sec:SR:Val}
 
% Spreading Rate comparisun in exp sim
Table~\ref{tab:SR} shows the validation comparison error, $e_{K_q}$  (Eq.~\ref{eq:D_one}) and the validation uncertainty, $\Delta_{K_q}$ from the experiment and the simulations. As discussed above, distributions of the model parameters of interest are generally assumed to come from symmetric distributions \cite{Maupin2018}. 
Because the spreading rates calculated from the sampling angle uncertainty in the \simDDD~simulation produced asymmetric bounds on $K_q$, two standard deviations of the distribution of the spreading rates was taken, producing $2\sigma_{u_{y}} = 0.021$ and $2\sigma_{\rho} = 0.017$. These standard deviations are large and dominate the validation uncertainty, once again, due to the grid imprints and under-sampling. Because of this, and computational resources, the validation uncertainty components for $\delta_{\mathrm{IC}}$ and $\delta_{\mathrm{Comp}}$ were set to $2\%$. The validity of these assumptions is discussed in Sec.~\ref{sec:Val}.

From Tab.~\ref{tab:SR}, the \simDDD~simulations $K_{u_{y}}$ ($K_{\rho}$) has a percent difference of $\approx 33\%$ ($\approx 45\%$) when compared to the experiment, whereas the \simDD~simulations $K_{u_{y}}$ ($K_{\rho}$) value has a percent difference of $\approx 310\%$ ($\approx 286\%$). The ratios of $e_{K_q}/\Delta_{K_q}$ for the \simDD~simulation show values much larger than $1$, indicating a systematic difference between the spreading rates. This is because the \simDD~simulation is a plan jet simulation, not a round jet. For $K_{u_y}$ in the \simDDD~simulation, $e_{K_{u_y}}/\Delta_{K_{u_y}}  = 0.71 < 1$, indicating an ambiguous result, due to the under-sampled simulation data leading to relatively large $\delta_{\mathrm{Num}}$. Ambiguous predictive accuracies are automatically within the the acceptance range, and must be discussed in the Validation Evaluation (Sec.~\ref{sec:Val}). For $K_\rho$, the ratio of validation comparison error to validation uncertainty does not show evidence of ambiguity or systematic errors, with a value of $e_{K_{u_\rho}}/\Delta_{K_{u_\rho}}  = 1.34$.

For the Model Acceptability criterion, we require $e_{K_q}$ to be less than the uncertainty in the experimental $K_q$'s. For the \simDDD~simulations $K_{\rho}$, the estimate of the Predictive Accuracy is $\delta_{K_{\rho}} = 0.024 \pm 0.018$ (Eq.~\ref{eq:PA}), producing a lower uncertainty range limit of $\delta_{K_{\rho}} = 0.024 - 0.018 = 0.006$, which is greater than the acceptable range of $0.004$. This result means that out of the simulated spreading rates that did not show ambiguity or systematic errors, the predictive accuracy of $K_{\rho}$ for the \simDDD~simulation is outside of the acceptability range, and is rejected.

%%%%%%%%%%%%%%%%%%%%%%%%%%%%%%%%%%%%%%%%%%%%
\subsubsection{Comparison to $K_q$'s in the literature}
\label{sec:SR:lit}

Although both simulated jets spread significantly faster than the experimental jet, for both simulations and the experiment, $K_{\rho} > K_{u_{y}}$, meaning that the \sfsix~jet spread further in $\rho$ then $u_{y}$. This behavior is consistent with the literature as concentration spreads faster than velocity \cite{Djeridane1996}. 

Jet spreading rates have been investigated by many authors for a variety of variable density jets with different density ratios, $s = \rho_{\mathrm{jet}}/\rho_{\infty}$. Figure~\ref{fig:CL_SRvsFW} shows selected spreading rates found in the literature for both experimental and simulated jet studies. Even with percent differences of $32\%$, the $K_{u_y}$'s found in the \simDDD~simulation are not unreasonable given the amount of variation in the measurements from different simulated jet studies. The \simDDD~simulation also shows the lowest $K_{u_y}$ values out of the simulation studies plotted in Fig.~\ref{fig:CL_SRvsFW}. The \simDD~simulations spreading rates are considerably larger than those measured from experiment, and found in the literature. The closest value being the high resolution round jet study from Magi \cite{Magi2001} of $K_{u_y} = 0.18$ for $s = 1.0$, in their RANS simulations using a $k-\epsilon$ turbulence model. The relatively high value of the \simDD~simulations $K_{u_y}$, and the relatively low value of the \simDDD~simulations $K_{u_y}$ show that, at least for the ILES's in this study, the computational dimension and resolution have a much larger influence on $K_q$ then density ratio.

%%%%%%%%%%%%%%%%%%%%%%%%%%%%%%%%%%%%%%%%%%%%
\subsubsection{$K_q$ behavior}
\label{sec:SR:Be}

\begin{figure}[ht!]
\centering
\includegraphics[width=0.95\linewidth,]{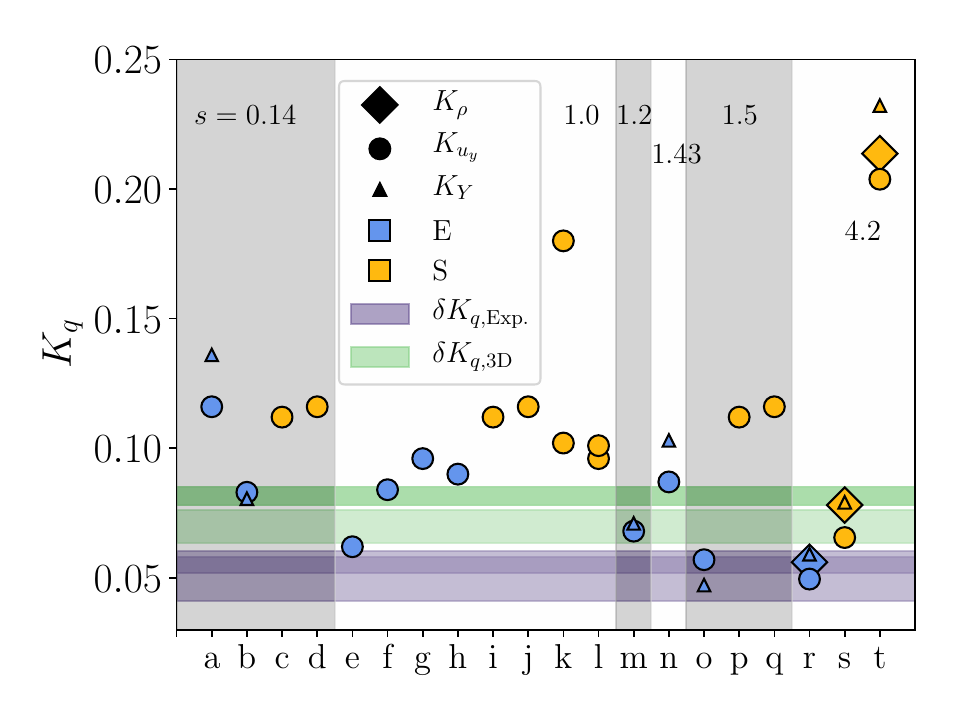}
\caption[]{Spreading rates, $K_{q}$, for mass fraction, downstream velocity and density from this work compared to those found in the literature for experimental (blue) and simulated (yellow) jet studies. The grey vertical shaded regions denote density ratio, the horizontal shaded regions denote the uncertainty in the TMT \sfsix~jet experiment (purple), and the \simDDD~simulation (green). The lettering denotes the studies where \textit{a} and \textit{g} are from Panchapakesan and Lumley \cite{PanchapakesanLumley1993}, \textit{b, e} and \textit{o} are from Djeridane et al. \cite{Djeridane1996}, \textit{c, i, p} and \textit{d, j, q} are the plane and round jet simulations of Foysi et al. \cite{Foysi2010}, \textit{f} is from Wygnanski and Fiedler \cite{Wygnanski1969}, \textit{h} and \textit{n} are from Chassaing et al. \cite{Chassaing1994}, \textit{k} and \textit{l} are the resolution study simulations of Magi \cite{Magi2001}, \textit{m} and \textit{r} are the air and \sfsix jets of Charonko and Prestridge \cite{Charonko2017}, \textit{s} and \textit{t} are the \simDDD, and \simDD~simulations.}
\label{fig:CL_SRvsFW}
\end{figure}

\begin{figure}[ht!]
\centering
\includegraphics[width=0.95\linewidth,]{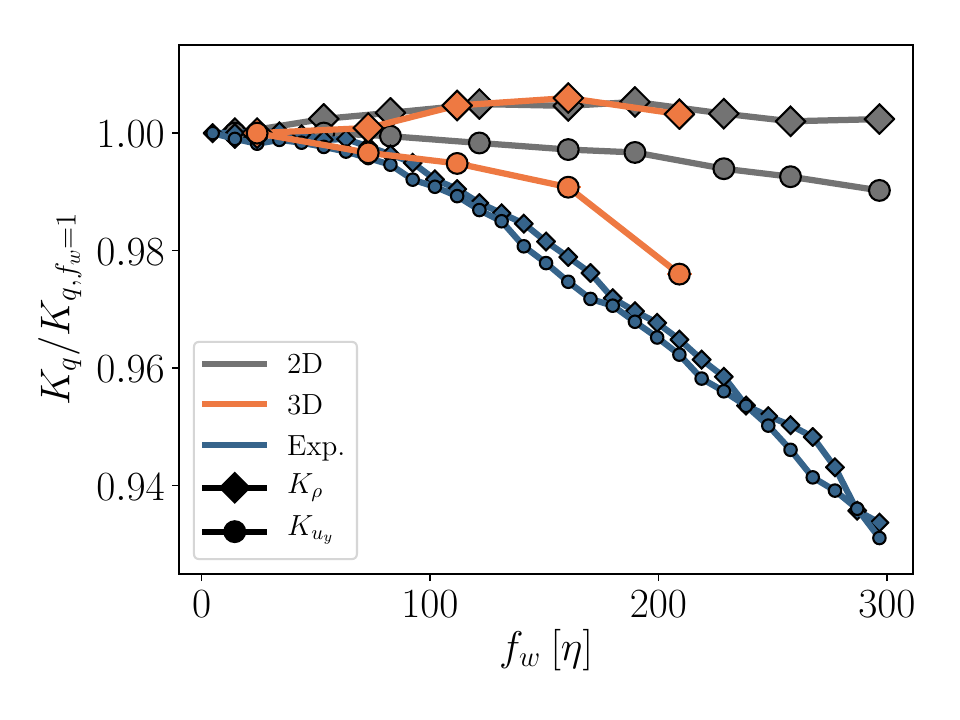}
\caption[]{Spreading rates for the experiment and the \simDD~and \simDDD~simulations vs filter width for both the $\rho$ and $u_{y}$ spreading rates. The spreading rates are scaled by the values measured on the grid scale.}
\label{fig:scaled_SR}
\end{figure}

% difference in SR with filter width
For any two flow variables, $q_1$ and $q_2$, it is not necessary that $K_{q_1} = K_{q_2}$, and in most cases, they are not the same (Fig.~\ref{fig:CL_SRvsFW}). Figure~\ref{fig:scaled_SR} shows $K_q$ scaled by the spreading rate measured at the grid scale (Tab.~\ref{tab:SR}) vs filter width for the experiment and both simulations. The filtering is done by a 2D Gaussian filter kernel (Sec.~\ref{sec:JetSimulations}) and can be thought of as a length scale decomposition. It is used here purely as an analysis tool to investigate the large scale structures of the jets on their respective grids. For $K_q$ in the experiment, as $f_w$ increases, both $K_{\rho}$ and $K_{u_{y}}$ decrease in a similar way. For the \simDD~simulation, both $K_{\rho}$ and $K_{u_{y}}$ remain roughly constant for increasing $f_{w}$. In the \simDDD~simulation, $K_{u_{y}}$ follows the decreasing trend of the experimental data, whereas $K_{\rho}$ remains constant similar to the \simDD~simulation. This result, as well as the larger percent difference when compared to the experiment ($45\%$), gives some evidence that in the \simDDD~simulation, the mixing of density is disproportionally larger than momentum in the variable density shearing layer, with the velocity growth being closer to reality.

%%%%%%%%%%%%%%%%%%%%%%%%%%%%%%%%%%%%%%%%%%%%
\subsection{Jet structure and mixing diagnostics}
\label{sec:jet_diag}
%%%%%%%%%%%%%%%%%%%%%%%%%%%%%%%%%%%%%%%%%%%%

This section is concerned with investigating differences in the behavior of the $\overline{u}_y$, $\overline{u}_r$ and $\overline{\rho}$ fields and providing some evidence as to why the the spreading rates found in the simulations are larger than in the experiment.  

 \subsubsection{Self similarity scaling}
 \label{sec:Self_Similarity_Scaling}

 % scale by what half width
Boussinesq jets in quiescent flow are self similar within the momentum-dominated regime. This means that the dynamics from different downstream locations scale. For variable density jets, as described by Charonko and Prestridge \cite{Charonko2017}, self similarity scaling for $\langle u_{y} \rangle$ and mass fraction is achieved by the radial similarity variable, $\eta_{r} = r/(y - y_{\widetilde{u}_{r}, 1/2})$ \cite{Chen1980, Charonko2017}, as well as the jet half-width $r_{q, 1/2}$. Scaling by $\eta_{r}$ illustrates the Gaussian nature of the downstream radial profiles and is able to show the difference in self similar states with respect to density ratio. Scaling by $r_{q, 1/2}$ is less subtle and collapses the profiles regardless of their centerline decay. 

\begin{figure}[ht!]
\centering
\includegraphics[width=0.95\linewidth,]{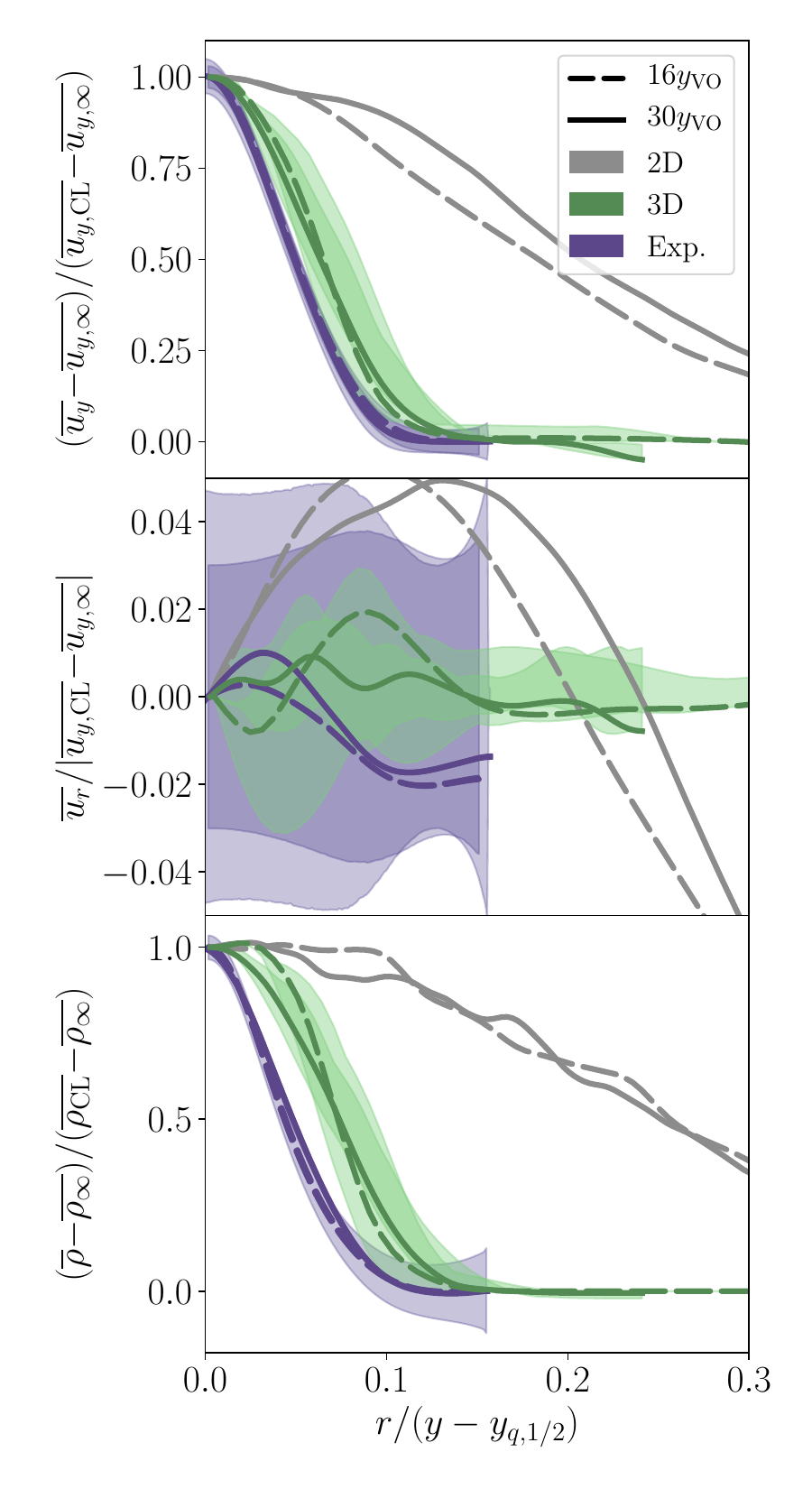}
\caption[]{Radially averaged profiles of the experimental (purple) and \simDDD~(green), and \simDD~(grey) simulation data showing the self-similarity variable, $\eta_{r}$, scaling. For $\overline{\rho}$, the virtual origin is taken as the density half-width virtual origin. For $\overline{u_{y}}$ and $\overline{u_{r}}$, the downstream velocity virtual origin is used. The y-axis values for $\overline{\rho}$ and $\overline{u}_{r}$, $\overline{u}_y$ are the scaled by the coflow subtracted center line values.}
\label{fig:self_similar_rprof_yq}
\end{figure}

Figure~\ref{fig:self_similar_rprof_yq} shows the scaled radial profiles for coflow subtracted $\overline{u_{y}}$, $\overline{u_{r}}$ and $\overline{\rho}$, scaled radially with the self-similarity variable, $\eta_{r}$. The larger jet spreading measured by the $K_q$ values found in the simulations is illustrated by these plots. The profiles for $\overline{u_{y}}$ and $\overline{\rho}$ are wider than the experimental profiles, implying larger spreading. The $\eta_{r}$ scaling scales the radial profiles to a self-similar state, although, as implied by the difference in spreading rates, this self-similar state is different for each case. The $\overline{u_{r}}$ profiles show significant deviation from the experimental profiles. The $\overline{u_{r}}$ profiles from the experiment are positive on the inside edge of the shearing region meaning jet gas is mixing into the shearing region, and become negative near the outside edge meaning the coflow air is being entrained. This behavior is not seen in the \simDDD~simulation $\overline{u_{r}}$ profiles, despite the growth of the shearing region, implying that there is too much variation in the $\overline{u_{r}}$ data to generate meaningful averages.

\begin{figure}[ht!]
\centering
\includegraphics[width=0.95\linewidth,]{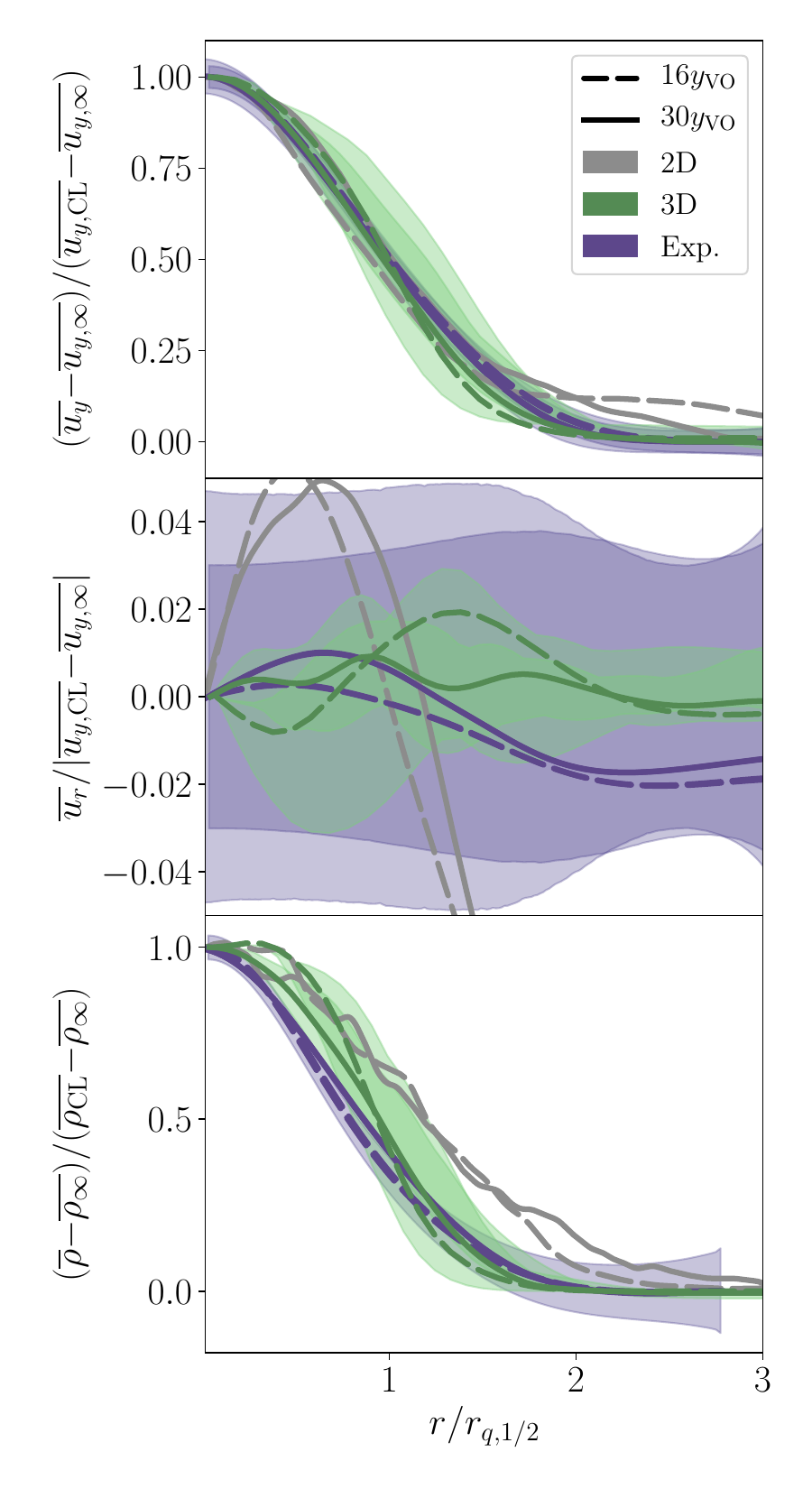}
\caption[]{Radially averaged profiles of the experimental (purple) and \simDDD~(green), and \simDD~(grey) simulation data showing the half-width self similarity scaling. The density is scaled radially by the local density half-width, $r_{\rho, 1/2}$, while $\overline{u}_y$ and $\overline{u}_{r}$ are scaled by the downstream velocity half-width, $r_{u_{y}, 1/2}$. The y axis values for $\overline{\rho}$ and $\overline{u}_{r}$, $\overline{u}_y$ are scaled by the coflow subtracted center line values.}
\label{fig:self_similar_rprof}
\end{figure}

Figure~\ref{fig:self_similar_rprof} shows the self similarity scaling using the jet half-width, $r_{q, 1/2}$. Along the y-axis, for density and downstream velocity, the profiles have been scaled by the coflow subtracted centerline values. The cross-stream velocity is also scaled by the coflow subtracted downstream velocity. Unlike the self-similarity variable scaling, the half-width scaling scales profiles regardless of differences in $K_q$. As Charonko and Prestridge \cite{Charonko2017} showed, the scaled experimental profiles for $\overline{u_{y}}$, $\overline{u_{r}}$ and $\overline{\rho}$ are self similar within the uncertainty. The scaled downstream velocities for the \simDDD~simulation also show self-similarity scaling within the sampling uncertainty with good agreement to the experimental profiles. For density, the kink in the centerline values can be seen in the profile for the \Bd~location (discussed in Sec.~\ref{sec:DSScaling}). For the \simDD~simulation, the scaled downstream velocity profiles seem to align with the experimental values for $r_{u_{y}, 1/2} < 1.5$ although are outside of the experimental uncertainty. For $r_{u_{y}, 1/2} > 1.5$, the scaled downstream velocities are faster than the experiment. Similarly, the scaled density profiles for the \simDD~simulation show considerable variation, and are outside of the uncertainty range of the experimental profiles. The $\overline{u_{r}}$ profiles once again show considerable variation for both simulations.

For both the radially self-similarity variable, and jet half-width scaling, uncertainty in the radial profiles is calculated as the relative uncertainty in the radial profiles. The relative uncertainty is calculated as the square root of the sum of squares of the measurement uncertainty in the fields over the image samples at a fixed radius.

The behavior of the scaled radial profiles, in general, reflect that of the spreading rates for the \simDDD~simulation, where quantities derived from $\overline{u_{y}}$ are closer to the experimental values then those derived from $\overline{\rho}$. 

%%%%%%%%%%%%%%%%%%%%%%%%%%%%%%%%%%%%%%%%%%%%
\subsubsection{Statistics and Variable-density Mixing}
\label{sec:Rprof_Stats}

% What is var, skw, krt
This section will examine the statistics of the radial profiles for $\overline{\rho}$, $\overline{u_r}$ and $\overline{u_y}$ in the free-shear layer. The intermittency in the shearing layer, and density field intermittency will be investigated and related to the spreading rates. 

The radial profiles of the skewness and kurtosis for $u_y$ and mass fraction have characteristic profiles for Boussinesq free-shear flows \cite[Sec. 5.5.3]{Wygnanski1969, Pope2000}. These skewness and kurtosis profiles diverge from near Gaussian values ($skw(u) = 0, \: krt(u) = 3$) near the centerline to larger values in the outer edges of the shearing region. The sharp increase in these values with respect to radius describes the turbulent free shear layer intermittence \cite{Zhou2018}, as eddies from within the mixing region extend outward into the coflow. For Boussinesq free-shear flows, these increases in skewness and kurtosis can be scaled to nearly Gaussian values by conditionally averaging with jet concentration \cite{LaRue1974}. In other words, the intermittent turbulent velocities characterized by these profiles mix fluid between the jet and coflow, and transport jet material from within the shearing region to the outer edge. The variance, skewness, and kurtosis were found for PDFs of the filtered flow variables, $\overline{\rho}$, $\overline{u}_r$ and $\overline{u}_y$ in the downstream homogenous direction for the $y_{u_{y}, 1/2}$ scaled \Bd~and \Cd~locations. For the data considered here, the variance of the sample is given by

\begin{equation}
\mathrm{var}(\overline{q}) = \langle {q^{\prime}}^2 \rangle_{b} = b(N) \sum_{i=0}^{N} (q(r, y_{i}) - \overline{q}(r))^2
\end{equation}

\noindent
where $q$ is the turbulent flow variable and $b(N) = 1/(N-1)$ is the Bessel corrected multiplicative factor, needed for the small sample counts of the simulations. The skewness and kurtosis are given by

\begin{equation}
\mathrm{skw}(\overline{q}) = \frac{\langle {q^{\prime}}^3 \rangle_{s}}{\langle {q^{\prime}}^2 \rangle_{s}^{3/2}}, \qquad
\mathrm{krt}(\overline{q}) = \frac{\langle {q^{\prime}}^4 \rangle_{s}}{\langle {q^{\prime}}^2 \rangle_{s}^2}
\end{equation}

\noindent
where the subscript $s$ denotes that the Bessel corrected multiplicative factor was replaced with $s(N) = 1/N$. It is important to note that the simulation data sets have a relatively low sample count compared to the number of experimental samples. This could potentially produce misleading results when dealing with higher order statistics. The remainder of this section focuses on the morphological differences between the experimental profiles and the \simDDD~simulation.

\begin{figure*}[ht!]
\centering
\begin{subfigure}{0.32\textwidth}
\includegraphics[width=\linewidth,]{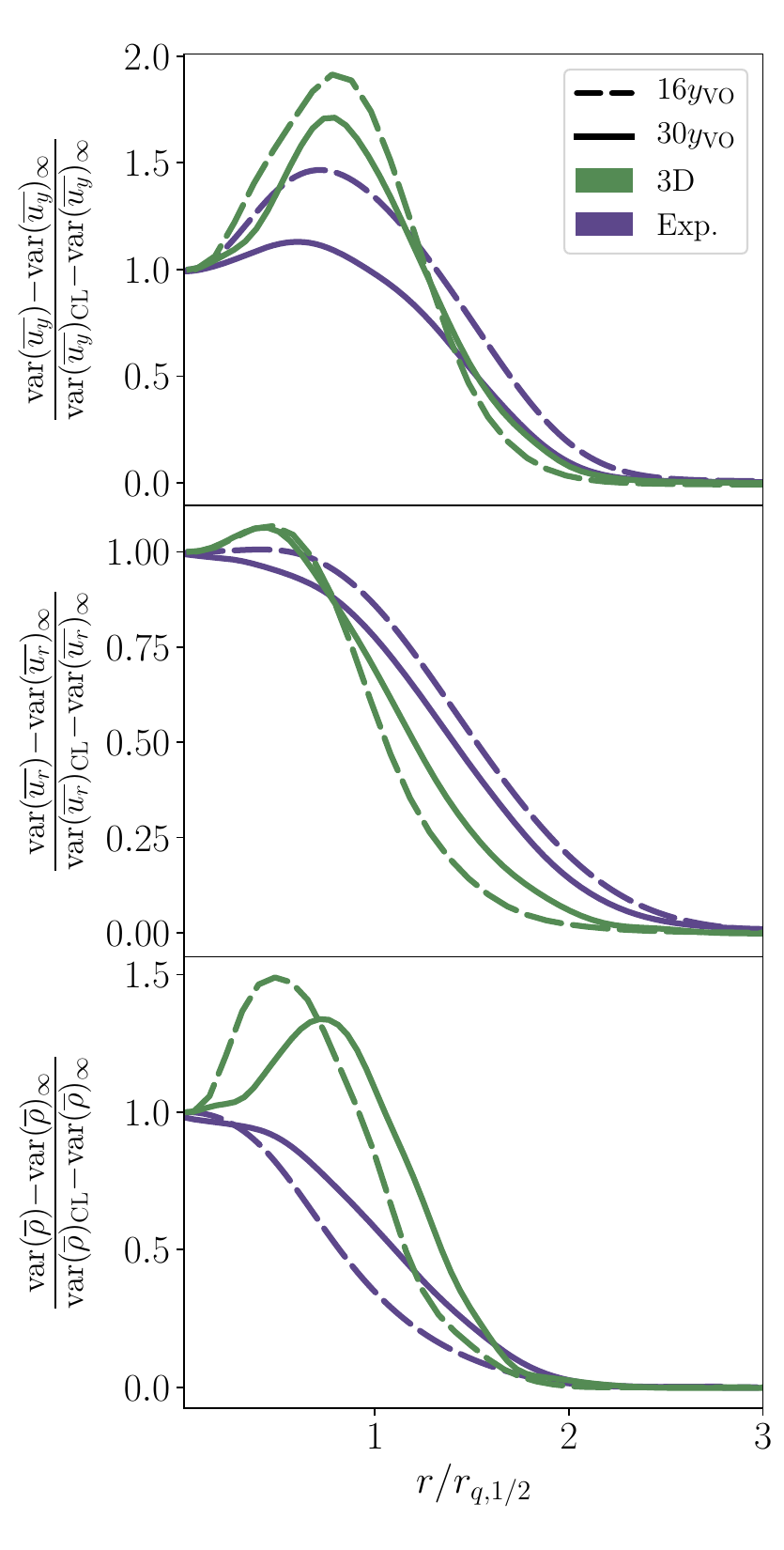}
\caption{Scaled radial variance profile} 
\label{fig:rprof_var}
\end{subfigure}
\begin{subfigure}{0.32\textwidth}
\includegraphics[width=\linewidth,]{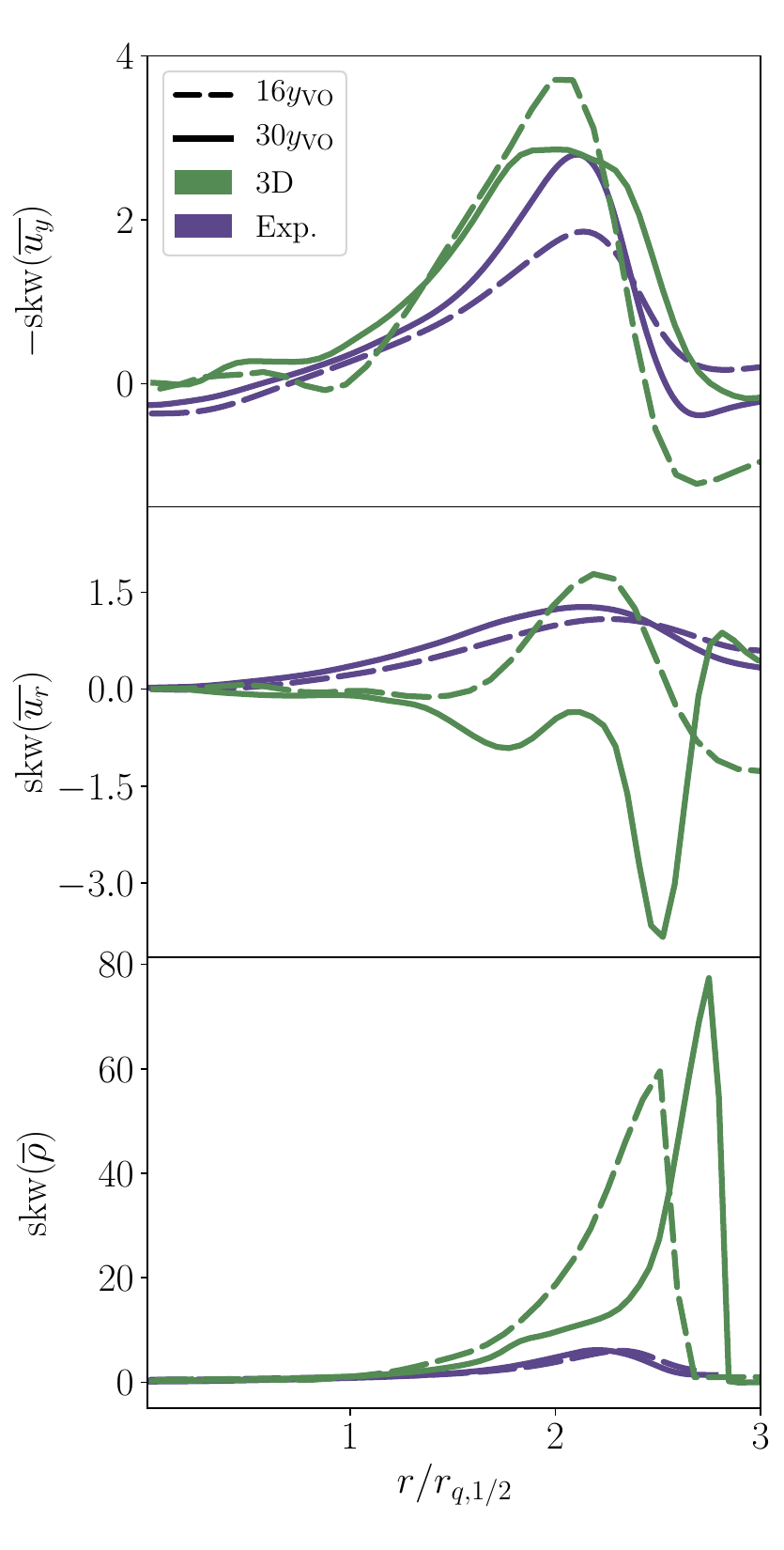}
\caption{Radial skewness profile} 
\label{fig:rprof_skw}
\end{subfigure}
\begin{subfigure}{0.32\textwidth}
\includegraphics[width=\linewidth,]{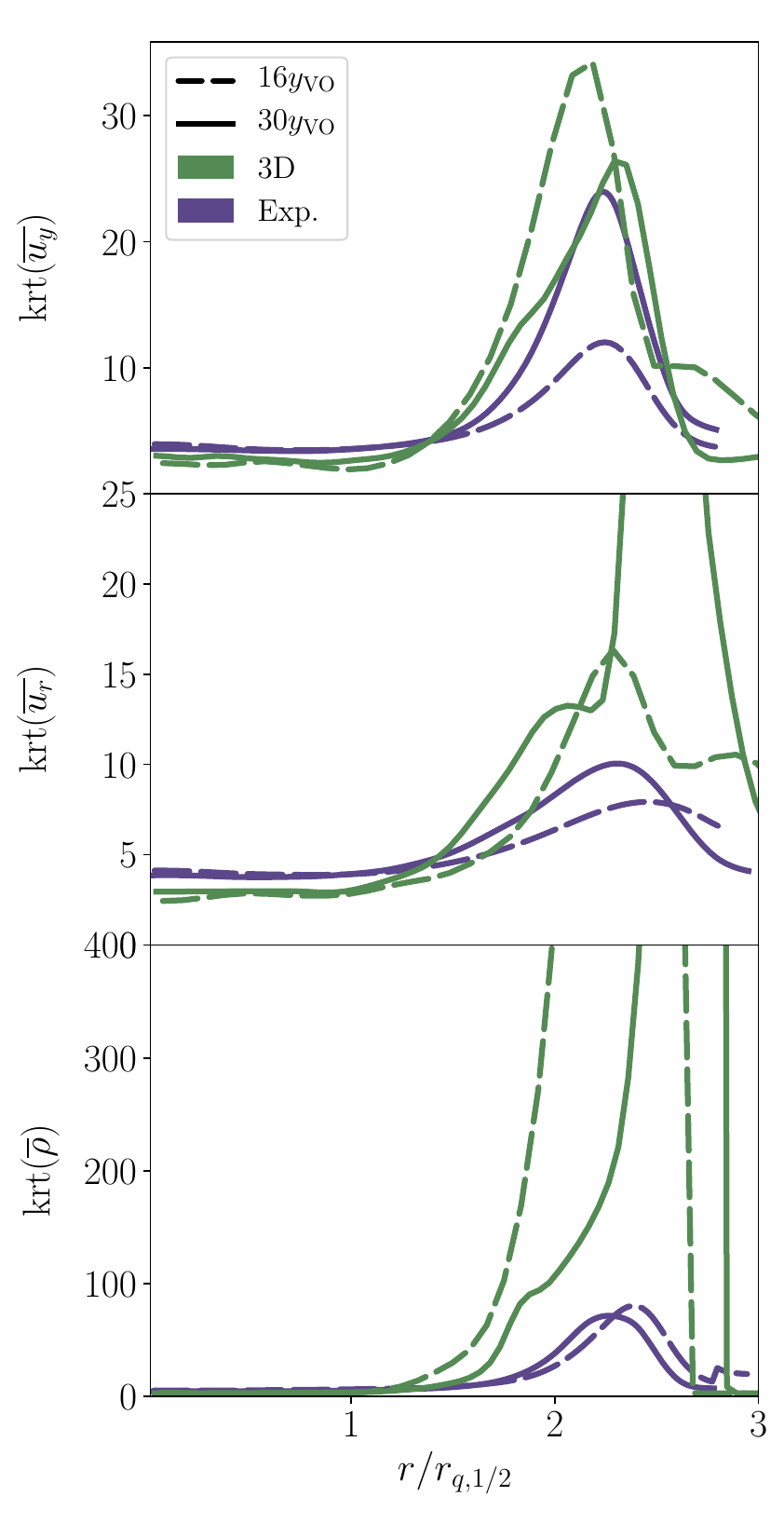}
\caption{Radial kurtosis profile} 
\label{fig:rprof_krt}
\end{subfigure}
\caption[]{Radially profiles of the variance (a), skewness (b), and kurtosis (c) for the experimental (purple), \simDDD~(green) simulation data. The radius is scaled using the self similar velocity halfwidth scaling as in Fig.~\ref{fig:self_similar_rprof}. The variance is scaled by the coflow subtracted centerline values.}
\label{fig:rprof_stats}
\end{figure*}

The panel plots for the variance, skewness, and kurtosis can be found in Fig.~\ref{fig:rprof_stats}. These plots have the same radial half-width self similarity scaling as Fig.~\ref{fig:self_similar_rprof}, so that the features of these statistical profiles can be compared to the mean radial profiles of the shearing region. The \simDD~simulation data was not used in this section as the statistical profiles are considerably noisy. 

From the skewness and kurtosis plots in Fig.~\ref{fig:rprof_skw}, \ref{fig:rprof_krt}, the experimental data shows the characteristic increase near the outer edge of the shearing region around $1.5r_{q, 1/2}$ in all quantities. The increased velocity intermittency near the edge of the shearing region corresponds to the increased density intermittence, specifically for $\overline{u_r}$, where the peaks in kurtosis are found at a similar radius (Fig.~\ref{fig:rprof_krt}, middle and bottom panel). Intermittent eddies from within the shearing region transport higher density material to the outer edge of the shearing region, further facilitating its growth. 

The \simDDD~simulation $\overline{u_y}$ statistics show a similar intermittency structure to those of the experiment (Fig.~\ref{fig:rprof_skw}, \ref{fig:rprof_krt}, top panel). Although the profiles do show differences, the overall morphology is similar, with the characteristic increase past $1.5r_{u_{y}, 1/2}$. The \simDDD~simulations $\mathrm{skw}(u_y)$ and $\mathrm{krt}(u_y)$ profiles are also larger than the experimental profiles past $1.5r_{u_{y}, 1/2}$, with the peak in kurtosis profile being about three times as large as in the experiment. The $\overline{u_r}$ kurtosis profile also shows some morphological similarity to the experimental data, at least in the \Bd~location. At the \Cd~location, past $2r_{u_{y}, 1/2}$ the kurtosis increases to very large values with respect to the experimental data (Fig.~\ref{fig:rprof_krt}, middle panel), which could be related to the variation seen in the skewness profile at the \Cd~location (Fig.~\ref{fig:rprof_skw}, middle panel). In general, the velocity intermittency characterized by the skewness and kurtosis profiles for the \simDDD~simulation is comparable to that of the experiment, although with larger values, meaning that the PDFs are more heavily skewed in downstream velocity, with narrower peaks and wider tails. This implies that the \simDDD~simulation experiences a higher probability of intermittent mixing events in the outer edge of the shearing region. 

The skewness and kurtosis for $\overline{\rho}$ in the \simDDD~simulation also show similar morphology to the experimental data, although the values of the peaks are considerably larger (Fig.~\ref{fig:rprof_skw} and Fig.~\ref{fig:rprof_krt}, bottom panel). Past $r_{u_{y}, 1/2}$ the skewness and kurtosis profiles of the \simDDD~simulation begin to increase. The kurtosis reaching a peak of $\approx 4000$ at \Cd, $200$ times larger than the experimental value, and a skewness peak of $\approx 80$ at \Cd, $100$ that of the experiment. The \simDDD~simulation experiences considerably higher density intermittency in the outer edge of the shearing region than the experiment, meaning there is a higher probability of higher density jet fluid being found at the outer edge of the shearing region, increasing the density mixing rates. 

Figure~\ref{fig:rprof_var} shows the coflow subtracted scaled variance profiles for $\overline{u_{y}}$, $\overline{u_{r}}$ and $\overline{\rho}$. The profiles are scaled to show the relative morphology. In general, the unscaled variance of the \simDDD~simulation profiles are larger than the experiment. In all profiles for the \simDDD~simulation, the variance has a peak between $0.5r_{u_{y}, 1/2}$ and $0.75r_{u_{y}, 1/2}$. This is in contrast to the experimental profiles which only have a peak in the variance of $\overline{u_y}$. In the simulation, the density variance is notable large and has a different morphology than the experimental profiles. This means that in the simulation, away from the centerline, there is a highly turbulent region with a reservoir of relatively unmixed jet gas, when compared to the experiment. If eddies from within this part of the shearing region are responsible for some contribution to the intermittency seen in the outer edge of the shearing region, they would bring with them relatively high density jet gas concentration.

The results of the statistical diagnostics that characterize the intermittency and turbulent mixing in the shearing layer are reflected in the spreading rates. Determination of the spreading rates showed that density spreads further than velocity in the \simDDD~simulation, and that the velocity spreading is larger than that in the experiment. This means that even with the relatively small number of slicing samples compared to the number of experimental images, the higher order statistical profiles seem to produce interpretations in agreement with the spreading rate behavior. Intermittent free-shear flow mixing events transport higher density material from within the shearing region further into the coflow. In the \simDDD~simulation, because there is higher intermittency in the density field within this shearing region, and the velocity structure of these mixing events are comparable to the experiment, dense fluid elements from within the shearing region are transported into the coflow more effectively, with less intermediate mixing, leading to increased mixing layer growth and therefore, larger $K_{\rho}$.

%%%%%%%%%%%%%%%%%%%%%%%%%%%%%%%%%%%%%%%%%%%%
\section{Discussion}
\label{sec:Disc}
%%%%%%%%%%%%%%%%%%%%%%%%%%%%%%%%%%%%%%%%%%%%

\subsection{Modeling uncertainties}
\label{sec:Dis:Mod}

% Inflow model
The inflow model assumes the PDFs of the velocity fluctuations in the coflow and within the inviscid core of the jet are Gaussian (Sec.~\ref{sec:inflow_model}). Although the fit to the coflow data in Fig.~\ref{fig:turb_model_PDFs} fit well around the means, turbulence in the jet core is not Gaussian because of the velocity shear. The inflow fluctuations are also not correlated in space and time, which can be problematic for turbulent inflow \cite{Klein2003}. Because the simulation requires computational time steps much smaller than the integral time (near which the experimental data is sampled), the simulations should see the inflow fluctuations as eddies with coherent structure in space and time. This data was not known from the experiment, so the inflow model prescribes fluctuations randomly. The effect of this on the flow would be to increase the turbulent kinetic energy at the inflow compared to the experiment. Although, as seen in the simulations, the coarse grid has the opposite effect as both simulated jets break up much slower than the experiment (Fig.~\ref{fig:072_067_density}). This implies that the increased turbulent kinetic energy on the grid at the inflow is transferred to larger scales, causing the increased downstream velocities seen in the simulations (Sec.~\ref{sec:DSScaling}, Fig.~\ref{fig:filtered_CL_profiles}).

% Undersampling of sim data
Both the \simDDD~and \simDD~simulations show signs of being under sampled in time (Fig.~\ref{fig:072_067_density}, Sec.~\ref{sec:DSScaling}). The \simDDD~simulation has prominent grid imprints in both $\overline{\rho}$ and $\overline{u}$ meaning that it is under resolved in space (Fig.~\ref{fig:halfwidth_slice}, Sec.~\ref{sec:Simulations}). At some length scale, all ILES will have grid imprints, but an ILES must have sufficient grid resolution to remove these grid imprints from the large-scale flows. The under-sampling in time is problematic as it convolves the natural statistical variation of the turbulence with the solution the code is producing. A larger sample size would average out the kink in the density profile (Sec.~\ref{sec:DSScaling}) and allow for better agreement. This is particularly difficult for ILES's as the time step is limited by the Courant-Friedrichs-Lewy (CFL) condition \cite{CFL1928}, and the integral time steps are separated by a considerable number of computational time steps.

\subsection{Validation Evaluation and Recommendations}
\label{sec:Val}

% Spreading rates
The jet spreading rates found in the \simDD~and \simDDD~simulations were larger than the experimental values of $K_{u_{y}} = 0.048 \pm 0.008$ and $K_{\rho} = 0.054 \pm 0.004$, having percent differences in $K_{u_{y}}$ ($K_{\rho}$) of $\approx 310\%$ ($\approx 286\%$) and $\approx 33\%$ ($\approx 45\%$) respectively (Sec.~\ref{sec:SR}). The spreading rates calculated from the \simDD~simulation showed systematic errors as determined by the $e_{K_q}/\Delta_{K_q}$ ratio of Wilson and Koskelo \cite{Wilson2018}. This result is consistent with other diagnostics of this study which show that the behavior of the simulated plane jets is fundamentally different from that of a round jet. For the \simDDD~simulation, the $e_{K_q}/\Delta_{K_q}$ ratio showed an ambiguous result for the $K_{u_y}$. Because the variable density jet is a multi-physics problem, it is necessary for both $K_{u_y}$ and $K_{\rho}$ to be within the Model Acceptability range for the model to be accepted. An ambiguous predictive accuracy is automatically within the Model Acceptability range. The ambiguity of the validation predictive accuracy indicates, in this case, that the validation uncertainty is far too large, being dominated by the variance of the simulation data. This variance is caused by two things, the low grid cell resolution leading to grid imprints, and a lack of samples, both of which increase the variation in the spreading rates of the \simDDD~simulation. Because of this large variance, a detailed analysis of the numerical uncertainties in the simulations was not carried out and is left for further work. A simulation with high enough grid cell resolution to eliminate the grid imprints would produce cylindrically symmetric velocity and density statistics. The set of spreading rates calculated from different angles of these cylindrically symmetric distributions could be considered estimated sample means of the same population, and would be normally distributed by the Central Limit Theorem. With azimuthal grid imprints the Central Limit Theorem does not apply. Because the $K_q$'s in the simulation were larger than the experiment with asymmetric distributions, assuming a symmetric PDF falsely increased the uncertainty in the simulations $K_q$'s. This leads to an overestimate in the validation uncertainty, and an underestimate in the $e_{K_q}/\Delta_{K_q}$ ratio. Because the simulations standard deviation dominates the validation uncertainty, an asymmetric uncertainty analysis would increase the $e_{K_q}/\Delta_{K_q}$ ratio, possibly to a point where this diagnostic reports the systematic grid imprint seen in the $u$ and $\rho$ fields.

An improvement on the Model Accuracy assessment method would be to downsample the experimental data to that of the simulation so that the variation in the parameters of interest are comparable. This was not done here as sample counts under 500 images for the experimental data produce prohibitively large variations in the radial profiles and spreading rates. With sufficient grid point resolution in the simulations and a high enough sample count, this Model Accuracy assessment method should produce reasonable and comparable results. Therefore we recommend that a higher resolution 3D simulation with sufficient time steps be calculated and reassessed given the appropriate methodology improvements above.

\subsection{Spreading rates}
\label{sec:dis:SR}

Despite being outside of the validation acceptable range, the \simDDD~simulations $K_{u_y} = 0.066$ is well within the relatively wide range of spreading rate values found in the experimental literature, and shows the smallest spreading rates out of the simulations (Sec.~\ref{sec:SR}, Fig.~\ref{fig:CL_SRvsFW}). The \simDD~simulation has the largest $K_{u_y}$ in Fig.~\ref{fig:CL_SRvsFW}. The simulated spreading rates from the literature plotted in Fig.~\ref{fig:CL_SRvsFW} use a variety of different computational methods (RANS, LES, ILES), are for a variety of jet types (plane and round), and use different or no models for the subgrid mixing ($k-\epsilon$, Smagorinsky). These studies are not necessarily comparable to each other but represent the possible range obtainable from different modeling assumptions. Although there may be a decreasing relation between spreading rate and density ratio in the experimental values, the variation in these values is still large. This trend is not seen in the simulated spreading rates most likely because the spreading rate values depend much more on the details of the computation rather than the density ratio.

\subsection{Jet scaling}
\label{sec:Dis:Scaling}

% VO scaling
For validation studies focused on the mixing properties of turbulent jets, $y_{q,1/2}$ scaling  worked well in finding comparable flow regions for the \simDDD~simulation (Sec.~\ref{sec:DSScaling}). Methods that target the scaling on the shearing region are particularly important for LES jet studies focussed on the downstream mixing as the effective sample size of $r_{q,1/2}$ can be increased by slicing or azimuthal averaging, and have less dependence on jet nozzle details. Other dynamical scalings that use dimensionless numbers (such as the downstream Froude number scaling of Chen and Rodi \cite{Chen1980}) may be more advanced for variable density jet studies, but these scalings only use the centerline profiles can suffer greatly from under-sampled data. 

\iffalse
% DS CL scaling
With simulations of sufficient resolution and number of samples, the downstream $\eta_r$ scaling could also be used to scale simulation and experimental jets with sufficiently large density ratio (Sec.~\ref{sec:CLScaling}). A potential problem when applying this scaling to ILES simulations is that it is valid in the near nozzle regions, so care must be taken to sufficiently resolve the nozzle and jet breakup.
\fi

\subsection{Mixing and jet structure}
\label{sec:Mixing}

% Stats 
Evidence of increased mixing across the shearing region in the \simDDD~simulation was found in the statistical profiles for $\overline{u}_{y}$, $\overline{u}_{r}$, and $\overline{\rho}$ (Sec.~\ref{sec:Rprof_Stats}). Despite the \simDDD~simulation having a relatively low sample count compared to the experimental data set, morphological similarities were found in the scaled higher-order statistics, namely, the skewness and kurtosis (Fig. \ref{fig:rprof_skw}, \ref{fig:rprof_krt}).  The statistical profiles of $\overline{u}_{y}$ in the \simDDD~simulation showed morphological similarities with the experimental profiles, indicating that the intermittency in the shearing region is comparable, although with marginally higher levels of turbulence. The profiles of $\mathrm{skw}(\rho)$ and $\mathrm{krt}(\rho)$ on the other hand were much larger than the experimental profiles, indicating a higher probability of less well mixed, higher density gas near the outer edge of the shearing region. The density variance profiles show a peak near the center of the shearing region, a feature that is not seen in the experimental density variance. This is representative of a highly turbulent region with a reservoir of relatively unmixed jet gas, at $r/r_{q,1/2} \approx 0.75$. If turbulent eddies from within this region of the shearing layer contribute to the intermittency near the outer edge, they will advect less mixed, higher density jet gas, increasing the density spreading. 

The transport of highly intermittent density fluctuations to the outer edge of the shearing region in the \simDDD~simulation explains the relatively large difference in the density spreading rates compared to velocity. Although both spreading rates were larger than the experiment, the $K_{\rho}$ is disproportionally larger than $K_{u_{y}}$. Because the statistical structure of the velocity is similar to the \simDDD~simulation and the experiment, but the density intermittency is larger, density spreads more creating a disproportionally large $K_{\rho}$.

%%%%%%%%%%%%%%%%%%%%%%%%%%%%%%%%%%
\subsection{Future Work}
\label{sec:FW}

An \sfsix~jet simulation run with a fine enough grid resolution to remove the grid imprints, and enough time steps to smooth the centerline profiles, would allow for a more detailed investigation of the variable density mixing in ILES. Higher quality simulation data would allow a validation study to move away from large-scale jet structure diagnostics to more physically motivated behavior governing the multi-physics aspects of variable density mixing layers such as analysis of vortex stretching \cite{Lai2018}. With a larger sample size, more advanced validation metrics could be applied. Metrics that treat the experimental measurements and model predictions as statistical distributions \cite{Maupin2018}. Results from a study such as this would be valuable in validating Reynolds averaged Navier-Stokes simulations of similar flow, as well as providing useful physical insight into multi-physics variable-density mixing in other flows using ILES.

%%%%%%%%%%%%%%%%%%%%%%%%%%%%%%%%%%%%%%%%%%%%
\section{Conclusion}
\label{sec:Conc}
%%%%%%%%%%%%%%%%%%%%%%%%%%%%%%%%%%%%%%%%%%%%

Although the validation study showed that higher quality simulation data is required for the model to be accepted, the methods used to perform the validation study could be easily applied to higher quality data. Methods like filtering to common grid scales and the downstream jet half-width scaling enforce the validity of further diagnostics by ensuring the comparison of similar turbulent flow regimes. This allows detailed validation analysis to more accurately determine regions of agreement and disagreement between the simulated and experimental jets.

%%%%%%%%%%%%%%%%%%%%%%%%%%%%%%%%%%%%%%%%%%%%%
\begin{acknowledgment}
AD would like to thank Patrick Payne for his work on coding the inflow model. 

This work was supported by the U.S. Department of Energy through the Los Alamos National Laboratory. Los Alamos National Laboratory is operated by Triad National Security, LLC, for the National Nuclear Security Administration of U.S. Department of Energy (Contract No. 89233218CNA000001).

\end{acknowledgment}

%%%%%%%%%%%%%%%%%%%%%%%%%%%%%%%%%%%%%%%%%%%%%

\begin{nomenclature}
\entry{$K_q$}{Jet spreading rate for the flow quantity, $q$}
\entry{$N$}{Number of samples}

\entry{$d_{0}$}{Jet nozzle diameter}
\entry{$dx$}{Grid spacing}
\entry{$e_{K_{q}}$}{Validation comparison error in $K_{q}$}
\entry{$f_{w}$}{Filter width}
\entry{$q$}{Flow quantity ($u_{y}$ or $\rho$)}
\entry{$<q>$}{Reynolds average of $q$}
\entry{$\overline{q}$}{Filtered average of $q$}
\entry{$q^{\prime}$}{Reynolds averaged fluctuation}
\entry{$r_{q, 1/2}$}{Half-width}
\entry{$s$}{Density ratio}
\entry{$u$}{Velocity}
\entry{$u_{y}$}{Downstream velocity}
\entry{$u_{r}$}{Radial velocity}
\entry{$u_{\mathrm{CL}}$}{Centerline velocity}
\entry{$u_{\mathrm{\infty}}$}{Coflow velocity}
\entry{$u_{\mathrm{jet}}$}{Downstream velocity at the nozzle}
\entry{$y_{q, 1/2}$}{Half-width virtual origin}
\entry{$y_{\mathrm{vo}}$}{Half-width shifted downstream region}

\entry{$\mathrm{var}(q)$}{Variance of $q$}
\entry{$\mathrm{skw}(q)$}{Skewness of $q$}
\entry{$\mathrm{krt}(q)$}{Kurtosis of $q$}

\entry{$\Delta_{K_{q}}$}{Validation uncertainty in $K_{q}$}
\entry{$\gamma$}{Adiabatic gas constant}
\entry{$\delta_{K_{q}}$}{Predictive accuracy}
\entry{$\delta_{\mathrm{Exp}}$}{Experimental uncertainty}
\entry{$\delta_{\mathrm{Num}}$}{Numerical uncertainty}
\entry{$\delta_{\mathrm{IC}}$}{Uncertainty in the simulation input}
\entry{$\delta_{\mathrm{Com}}$}{Comparison uncertainty}
\entry{$\eta$}{Kolmogorov length scale}
\entry{$\theta$}{Azimuthal angle in cylindrical coordinates}
\entry{$\rho$}{Density}
\entry{$\rho_{\mathrm{CL}}$}{Centerline density}
\entry{$\rho_{\mathrm{\infty}}$}{Coflow density}
\entry{$\rho_{\mathrm{jet}}$}{Density of pure jet gas}
\entry{$\sigma$}{Standard deviation}

\end{nomenclature}

%%%%%%%%%%%%%%%%%%%%%%%%%%%%%%%%%%%%%%%%%%%%%
\bibliographystyle{asmems4}
\bibliography{refs}

\end{document}